\documentclass[11pt]{article}%
\usepackage{amsmath}
\usepackage{graphicx}
\usepackage{array}
\usepackage{booktabs}
\usepackage{subfigure}
\usepackage{float}
\usepackage{url}
\usepackage{color}
\usepackage[ruled,linesnumbered]{algorithm2e}%
\usepackage{amsfonts}%
\usepackage{amssymb}
\newcommand{\eqnb}{\begin{equation}}
\newcommand{\eqne}{\end{equation}}

\newtheorem{The}{Theorem}

\newtheorem{Rem}{Remark}

\begin{document}

\title{Dynamic Practical Byzantine Fault Tolerance and Its Blockchain System: A
Large-Scale Markov Modeling}
\author{Yan-Xia Chang$^{1}$, Quan-Lin Li$^{1}$, Qing Wang$^{2}$ \thanks{Corresponding
author: Q. Wang (qing.wang1@monash.edu)}, Xing-Shuo Song$^{3}$\\$^{1}$School of Economics and Management\\Beijing University of Technology, Beijing 100124, China\\$^{2}$Monash Business School, Monash University\\900 Dandenong Road Caulfield East, 3145, VIC, Australia\\$^{3}$School of Economics and Management\\Yanshan University, Qinhuangdao 066004, China}
\maketitle
\footnotetext{ This work has been submitted to the IEEE for possible publication. Copyright may be transferred without notice, after which this version may no longer be accessible.  }

\begin{abstract}
In a practical Byzantine fault tolerance (PBFT) blockchain network, the voting
nodes may always leave the network while some new nodes can also enter the
network, thus the number of voting nodes is constantly changing. Such a new
PBFT with dynamic nodes is called a dynamic PBFT. Clearly, the dynamic PBFT
can more strongly support the decentralization and distributed structure of
blockchain. However, analyzing dynamic PBFT blockchain systems will become
more interesting and challenging.

In this paper, we propose a large-scale Markov modeling technique to analyze
the dynamic PBFT voting processes and its dynamic PBFT blockchain system. To
this end, we set up a large-scale Markov process (and further a
multi-dimensional Quasi-Birth-and-Death (QBD) process) and provide performance
analysis for both the dynamic PBFT voting processes and the dynamic PBFT
blockchain system. In particular, we obtain an effective computational method
for the throughput of the complicated dynamic PBFT blockchain system. Finally,
we use numerical examples to check the validity of our theoretical results and
indicate how some key system parameters influence the performance measures of
the dynamic PBFT voting processes and of the dynamic PBFT blockchain system.
Therefore, by using the theory of multi-dimensional QBD processes and the
RG-factorization technique, we hope that the methodology and results developed
in this paper shed light on the study of dynamic PBFT blockchain systems such
that a series of promising research can be developed potentially.

\vskip           0.5cm


\end{abstract}

\section{Introduction}

Blockchain technologies originated in Bitcoin by Nakamoto \cite{Nak:2008} in
2008. Since then, Blockchain has attracted tremendous attention from both
research communities and industrial applications. Furthermore, many real
applications of blockchain benefit from a number of salient and excellent
features, for example, decentralization, distributed structure, availability,
persistency, consistency, anonymity, immutability, auditability, and
accountability. So far, blockchain has been envisioned as a powerful
backbone/framework for decentralized data processing and data-driven
autonomous organization in a peer-to-peer and open-access network. Readers may
refer to books by Narayanan et al. \cite{Nar:2016}, Bashir \cite{Bas:2018},
Raj \cite{Raj:2019}, Maleh et al. \cite{Mal:2020}, Rehan and Rehmani
\cite{Reh:2020} and Schar and Berentsen \cite{Sch:2020}; and survey papers by
Wang et al. \cite{Wan:2019}, Gorkhali et al. \cite{Gor:2020}, Belchior et al.
\cite{Bel:2021} and Huang et al. \cite{Hua:2021}; and further survey papers
with serval real areas by Fauziah et al. \cite{Fau:2020} for smart contracts,
Dai et al. \cite{Dai:2020} for Internet of Things (IoT), Sharma et al.
\cite{Sha:2020} for cloud computing, Gorbunova et al. \cite{Gor:2022} for
industrial applications, and Ekramifard et al. \cite{Ekr:2020} for artificial
intelligence (AI).

Consensus mechanisms always play a pivotal role in developing blockchain
technologies. Up to now, there have been more than 50 different consensus
mechanisms in the study of blockchain technologies. We refer readers to recent
survey papers by, for example, Cachin and Vukoli\'{c} \cite{Cac:2017}, Bano et
al. \cite{Ban:2017}, Natoli et al. \cite{Nat:2017}, Chaudhry and Yousaf
\cite{Cha:2018}, Nguyen and Kim \cite{Ngu:2018}, Ongaro and Ousterhout
\cite{Ongaro:2014}, Salimitari and Chatterjee \cite{Sal:2018}, Wang et al.
\cite{Wan:2019}, Pahlajani et al. \cite{Pah:2019}, Nguyen et al.
\cite{Ngu:2019}, Carrara et al. \cite{Car:2020}, Wan et al. \cite{Wan:2020},
Xiao et al. \cite{Xia:2020}, Ferdous et al. \cite{Fer:2020}, Nijsse and
Litchfield \cite{Nij:2020}, Leonardos et al. \cite{Leo:2020}, Yao et al.
\cite{Yao:2021}, Lashkari and Musilek \cite{Las:2021}, Fu et al.
\cite{Fu:2021}, Khamar and Patel \cite{Kha:2021}, Oyinloye et al.
\cite{Oyi:2021}, Bains \cite{Bai:2022} and Xiong et al. \cite{Xio:2022}.

For a reliable distributed computer system, the consensus result of its
components reaching an agreement on a certain state is the most fundamental
and important issue. To achieve consistency, a reliable distributed computer
system must be able to cope with the failure of one or more of its components,
in case a failed component can send conflicting information to different parts
of the computer system. To solve the type of failure and conflicting problems,
an important concept: \textit{Byzantine generals problem}, is developed, e.g.,
see Lamport et al. \cite{Lam:1982}, Lamport \cite{Lam:1983}, Schlichting and
Schneider \cite{Schlichting:1983}, Reischuk \cite{Rei:1985} and Martin and
Alvisi \cite{Mar:2016} for more details. Based on the Byzantine generals
problem, Pease et al. \cite{Pease:1980} and Lamport et al. \cite{Lam:1982}
proposed the Byzantine fault tolerant consensus mechanism (BFT), and further
research includes Thai et al. \cite{Thai:2019}, Li et al. \cite{LiY:2021},
Zhan et al. \cite{Zhan:2021} and so on.

Unfortunately, the original BFT has the problems of low algorithm efficiency,
small node capacity and weak scalability. To solve these problems, Castro and
Liskov \cite{Castro:1999} improve the BFT and proposed the PBFT (Practical
Byzantine Fault Tolerance consensus mechanism), which makes the BFT feasible
in many practical applications. Thereafter, some researchers further developed
the BFT to improve the performance of the BFT or PBFT effectively. Important
examples include Castro and Liskov \cite{Cas:2002}, Veronese et al.
\cite{Veronese:2011}, Abraham et al. \cite{Abr:2017}, Hao et al.
\cite{Hao:2018}, Gueta et al. \cite{Gueta:2019}, Malkhi et al.
\cite{Malkhi:2019}, Sakho et al. \cite{Sak:2020}, Nischwitz et al.
\cite{Nis:2021}, Oliveira et al. \cite{Oli:2022} and so on. Up to now, the BFT
and PBFT have become the most basic ones in all the blockchain consensus
mechanisms, and both play a crucial role in extending, generalizing, and
finding new effective blockchain consensus mechanisms. On the research line,
noteworthy examples include Kiayias and Russell \cite{Kiayias:2018}, Bravo et
al. \cite{Bra:2020}, Meshcheryakov et al. \cite{Mes:2021}, Alqahtani and
Demirbas \cite{Alq:2021}, Ma et al. \cite{Ma:2022}, Garcia et al.
\cite{Gar:2022}, Navaroj et al. \cite{Nav:2022} and so forth.

Different from those works in the literature, a key purpose of this paper is
to further propose and develop a new PBFT consensus mechanism in blockchain
technologies, called a dynamic PBFT consensus, in which the votable nodes may
always leave the PBFT network while some new nodes can also enter the PBFT
network. In this case, the number of votable nodes is constantly changing,
thus analysis of the dynamic PBFT is more challenging due to at least three
reasons as follows:

\textbf{(a) }Note that the votable nodes may always leave the PBFT network
randomly, the total number of votable nodes may become so small that the
votable nodes cannot represent the legitimacy of final vote results in the
PBFT network. Thus it highlights the need to establish a lower bound on the
total number of votable nodes to ensure that the legally voting process of the
PBFT network can be executed.

\textbf{(b)} Some new nodes can randomly enter the PBFT network, which further
underpins the decentralization and distributed structure of blockchain in a
huge P2P network range. Therefore, it is obviously inappropriate to design a
fixed number of voting nodes in the PBFT network. In addition, too many nodes
entering the network will exceed the capacity of the network, so setting up an
upper bound to realize the voting processes smoothly is a requisite in the
PBFT network.

\textbf{(c)} The major node and slave nodes deal with each transaction package
through three stages of parallel voting processes: Prepare, commit, and reply.
Thus, it is always challenging and complex to analyze such three-phase
parallel PBFT voting processes. See Ma et al. \cite{Ma:2022} for more details.

Based on the above analysis, it is important to study the dynamic PBFT voting
processes, and to provide performance evaluation of the dynamic PBFT
blockchain system. To this end, we propose a large-scale Markov modeling
technique to analyze the dynamic PBFT voting processes and its dynamic PBFT
blockchain system. We first set up a large-scale Markov process whose elements
are given a detailed discussion related to the dynamic PBFT. Then we provide
key performance measures of the dynamic PBFT voting processes. Furthermore, we
construct an approximate queueing model to discuss the dynamic PBFT blockchain
system and provide its performance analysis. It is worthwhile to note that we
provide a novel method to compute the throughput of the dynamic PBFT
blockchain system. Finally, we use numerical examples to verify the validity
of our theoretical results.

Note that Hao et al. \cite{Hao:2018}, Ma et al. \cite{Ma:2022}, and Nischwitz
et al. \cite{Nis:2021} are three closely related works to our paper. Hao et
al. \cite{Hao:2018} presented the dynamic PBFT network in which some nodes may
enter or leave the PBFT network by means of the consensus protocols: Using the
JOIN and EXIT protocols leads to some dynamic nodes. It is worthwhile to note
that Hao et al. \cite{Hao:2018} is different from our work given in this
paper, we describe and analyze the dynamic (entering and leaving) behavior of
some nodes in the PBFT network through using the Markov process theory or
random dynamical system. Ma et al. \cite{Ma:2022} considered a special case of
this paper (i.e., the voters are fixed) by means of a two-dimensional Markov
process. Nischwitz et al. \cite{Nis:2021} introduced a probabilistic model for
evaluating BFT protocols in the presence of dynamic link and crash failures.
Their analysis is different from our large-scale Markov modeling technique
developed in this paper, we observe the dynamic behavior of some nodes and
provide performance evaluation of the dynamic PBFT blockchain system by means
of the Markov process theory. By comparing the two studies, it is easy to see
that our large-scale Markov modeling technique is superior to their
probabilistic analysis method not only from the dynamic systems but also from
the performance evaluation.

The Markov processes and queueing theory play a key role in the study of
blockchain systems. Readers can refer to survey papers by, for example,
Smetanin et al. \cite{Sme:2020}, Fan et al. \cite{Fan:2020}, and Huang et al.
\cite{Hua:2021}. Up to now, some papers have applied the Markov processes (or
Markov chains) to study the blockchain systems. For example, a
transition-construction Markov chain by Eyal and Sirer \cite{Eyal:2014},
Markov queueing models by Li et al. \cite{Li:2018, Li:2019}, a two-dimensional
Markov process by G\"{o}bel et al. \cite{Goebel:2016}, a new computational
method further developed by Javier and Fralix \cite{Jav:2020}, a pyramid
Markov processes by Li et al. \cite{Li:2020}, and a Markov process of
DAG-based blockchain systems by Song et al. \cite{Song:2022}.

Based on the above analysis, the main contributions of this paper are
summarized as follows:

\begin{itemize}
\item[1.] This paper proposes a novel dynamic PBFT, where the votable nodes
may always leave the network while new nodes may also enter the network, thus
the number of votable nodes is constantly changing. Compared with the ordinary
PBFT, the analysis of the dynamic PBFT is more interesting and challenging. To
do this, we propose a large-scale Markov modeling technique to analyze the
dynamic PBFT voting processes and dynamic PBFT blockchain system.

\item[2.] For the dynamic PBFT voting processes, we set up a large-scale QBD
process and obtain its stationary probability vector, which is used to
numerically compute performance measures of the dynamic PBFT voting processes.
Accordingly, we establish an approximate queueing model to discuss the dynamic
PBFT blockchain system and provide a new method to analyze performance of the
dynamic PBFT blockchain system.

\item[3.] We use numerical examples to check the validity of our theoretical
results and indicate how some key system parameters influence performance
measures of the dynamic PBFT voting processes and the dynamic PBFT blockchain system.
\end{itemize}

The rest of this paper is organized as follows. Section \ref{sec:model
descrition} describes stochastic models for the dynamic PBFT voting processes
and the dynamic PBFT blockchain system. Section \ref{sec:voting process} sets
up a large-scale QBD process to express the dynamic PBFT voting processes.
Section \ref{sec:performnace measures} obtains the stationary probability
vector of the large-scale QBD process and provides performance measures of the
dynamic PBFT voting processes. Section \ref{sec:dynamic system} establishes an
approximate queueing model to discuss the dynamic PBFT blockchain system and
provides a new method to compute the throughput of the dynamic PBFT blockchain
system. Section \ref{sec:algorithms} provides two effective algorithms for
computing the throughput of the dynamic PBFT blockchain system. Section
\ref{sec:numberical analysis} uses some numerical examples to verify the
validity of our theoretical results and demonstrates how the performance
measures are influenced by some key system parameters. Some concluding remarks
are given in Section \ref{sec:concluding}.

\section{Model Description of the Dynamic PBFT}

\label{sec:model descrition}

In this section, we provide a detailed model description for the dynamic PBFT
with entering and leaving nodes. Furthermore, we give mathematical notation,
random factors, and necessary parameters used in our subsequent study.

In a dynamic PBFT, some nodes can always enter and leave the PBFT network. In
this situation, the number of votable nodes may be unfixed. Therefore, how to
describe and study such a dynamic PBFT becomes more interesting and challenging.

Now, we describe the dynamic PBFT with entering and leaving nodes as follows:

\textbf{(1) Nodes enter the PBFT network: }We assume that some external nodes
entering the PBFT network follow a Poisson process with arrival rate $\mu>0$.
Obviously, the newly entering nodes increase the number of votable notes, so
that the number of over two-thirds valid votes will also increase.

\textbf{(2) Nodes leave the PBFT network: }We assume that the time of each
valid voting node spent in the PBFT network is exponential with mean
$1/\theta>0$. Such a random time indicates that all nodes have an impatient
behavior that results from multiple reasons. For example, some nodes suddenly
go offline, some nodes change interest in participating in such voting, some
nodes are forcibly removed from the PBFT network, and so forth.

\textbf{(3) A lower threshold is required for the minimal number of valid
voting nodes: }Because the number of votable nodes changes randomly, we must
require a lower threshold for the minimal number of votable nodes. Such a
lower threshold is used to guarantee the security of the dynamic PBFT voting
process, that is, the dynamic PBFT voting process must have a sufficient
number of nodes to vote and reach a consensus. We assume that the lower
threshold is $\mathcal{M}$. If over $\mathcal{M}$ nodes vote and reach a
consensus, then the dynamic PBFT voting process is legal so that the voting
results can be accepted.

\textbf{(4) An upper threshold for the maximal number of valid voting nodes:}
For the convenience of analysis, we set an upper threshold for the maximum
number of valid voting nodes to avoid the infinite expansion of PBFT network
size when the external nodes constantly enter. Meanwhile, our purpose is to
avoid some complicated theoretical discussion for a large-scale Markov model
of the dynamic PBFT voting process, for example, stability analysis, and
computation of the stationary probability vector. We assume that the upper
threshold is $\mathcal{N}$. When the number of valid voting nodes reaches the
upper threshold $\mathcal{N}$, any new arriving external node can no longer
enter the PBFT network.

\textbf{(5) The probability that the transaction package is approved or
refused by the valid voting nodes: }To simplify the analysis, we assume that
all valid voting nodes are identical when a transaction package is submitted
to each node for voting. In this paper, we do not distinguish the properties
of valid voting nodes, such as Byzantine or non-Byzantine. Furthermore, we
assume that the voting time of each node is exponential with mean ${1}%
/{\gamma>0}$; and the probability that a transaction package is approved by
each node is $p$, while the probability that a transaction package is refused
by each node is $q=1-p$.

\textbf{(6) The judgment of the voting result: }We denote by $N(t)$, $M(t)$,
and $K(t)$ the number of valid voting nodes, the number of nodes that approve
the transaction package, and the number of nodes that refuse the transaction
package at time $t$, respectively. $N(t)-M(t)-K(t)$ is the number of nodes
that have not completed their voting processes yet. We assume that
\textbf{(a)} a transaction package becomes a block if $M(t)>\left(
2/3\right)  \cdot N(t)$ and $M(t)\geq\mathcal{M}$; and \textbf{(b)} the
transaction package becomes an orphan block if $M(t)\leq\left(  2/3\right)
\cdot N(t)$ and $M(t)\geq\mathcal{M}$, and it is returned to the transaction
pool (i.e., rollback).

\textbf{(7) The times of block-pegging and rolling-back: }The block-pegging
time is a time interval from the completion time of the voting and consensus
to the epoch that the block is pegged on the blockchain. Also, the
rolling-back time is also a time interval from the completion time of voting
and consensus to the epoch that the orphan block is returned to the
transaction pool.

Note that the times of block-pegging and rolling-back are mainly determined by
the network latency of the dynamic PBFT system, both of them are identical. In
this case, we assume that the block-pegging time and rolling-back time are
exponential with the same mean ${1}/{\beta}$.

\textbf{(8) Transaction arrivals at the transaction pool:} To study the
dynamic PBFT blockchain system (see Section \ref{sec:dynamic system}), we
assume that arrivals of transactions follow a Poisson process with arrival
rate $\lambda>0$, and the capacity of the transaction pool is infinite.

\textbf{(9) Independence: }We assume that all random variables defined above
are independent of each other.

\begin{Rem}
When $\mathcal{M}\leq N(t)\leq\mathcal{N}$, for each positive integer $N(t)$,
there exist a positive integer $k$ such that $N(t)=3k,3k+1,3k+2$. In fact, we
can find that checking (a) and (b) in Assumption (6) is not easy for some
integers, and it is necessary and useful for considering the following cases:
\textbf{(1)} If $N(t)=3k$ or $N(t)=3k+1$, $M(t)\geq2k+1$; and \textbf{(2)} If
$N(t)=3k+2$, $M(t)\geq2k+2$. Based on this, we write%
\begin{equation}
m_{l}=\left\{
\begin{array}
[c]{ll}%
2k+1, & N(t)=3k \text{ or } N(t)=3k+1,\\
2k+2, & N(t)=3k+2.
\end{array}
\right.
\end{equation}
\begin{equation}
k_{l}=\left\{
\begin{array}
[c]{ll}%
k, & N(t)=3k,\\
k+1, & N(t)=3k+1\text{ or }N(t)=3k+2.
\end{array}
\right.
\end{equation}
If $M(t)\geq m_{l}$, then the transaction package becomes a block; and if
$K(t)\geq k_{l}$, then the transaction package becomes an orphan block, which
is returned to the transaction pool.
\end{Rem}

\begin{Rem}
In our dynamic PBFT voting process, we describe the voting behavior of
Byzantine nodes from a probabilistic perspective, which is reasonable by means
of a statistical approach.
\end{Rem}

\section{A QBD Process for Dynamic PBFT Voting Process}

\label{sec:voting process}

In this section, we set up a three-dimensional continuous-time Markov model to
analyze the dynamic PBFT voting process, and further formulate it as a QBD
process with finite states.

Note that $N(t)$, $M(t)$, and $K(t)$ denote the number of valid voting nodes,
the number of valid voting nodes that approve the transaction package, and the
number of valid voting nodes that refuse the transaction package at time $t$,
respectively; and $N(t)-M(t)-K(t)$ is the number of valid voting nodes that
have not completed their voting process yet.

For convenience of analysis, we take that $\mathcal{M}=3L$ and $\mathcal{N}%
=3N+2$, where $L$ and $N$ are two fixed constants, and $N\gg L$.

It is clear that $\left\{  \left(  N(t),M(t),K(t)\right)  :t\geq0\right\}  $
is a three-dimensional continuous-time Markov process, whose state space is
given by
\begin{equation}
\Theta=\left\{  {(}l,0,0{):}0\leq l\leq3L-1\right\}  \cup\left\{
\bigcup\limits_{k=L}^{N}{{\mathrm{{Level}}\;}}k\right\}  ,
\end{equation}
where
\[
\mathrm{{Level}}\;k=\text{Sublevel}_{k,0}\cup\text{Sublevel}_{k,1}%
\cup\text{Sublevel}_{k,2},
\]%
\begin{align*}
\text{Sublevel}_{k,0}=  &  \{(3k,0,0),(3k,0,1),\ldots
,(3k,0,3k-2),(3k,0,3k-1),(3k,0,3k);\\
&  (3k,1,0),(3k,1,1),\ldots,(3k,1,3k-2),(3k,1,3k-1);\\
&  (3k,2,0),(3k,2,1),\ldots,(3k,2,3k-2);\ldots;\\
&  (3k,3k,0)\},
\end{align*}%
\begin{align*}
\text{Sublevel}_{k,1}=  &  \{(3k+1,0,0),(3k+1,0,1),\ldots
,(3k+1,0,3k),(3k+1,0,3k+1);\\
&  (3k+1,1,0),(3k+1,1,1),\ldots,(3k+1,1,3k-1),(3k+1,1,3k);\\
&  (3k+1,2,0),(3k+1,2,1),\ldots,(3k+1,2,3k-1);\ldots;\\
&  (3k+1,3k+1,0)\},
\end{align*}%
\begin{align*}
\text{Sublevel}_{k,2}=  &  \{(3k+2,0,0),(3k+2,0,1),\ldots
,(3k+2,0,3k+1),(3k+2,0,3k+2);\\
&  (3k+2,1,0),(3k+2,1,1),\ldots,(3k+2,1,3k),(3k+2,1,3k+1);\\
&  (3k+2,2,0),(3k+2,2,1),\ldots,(3k+2,2,3k);\ldots;\\
&  (3k+2,3k+2,0)\}.
\end{align*}

\begin{figure}[ptbh]
\centering          \includegraphics[width=7cm]{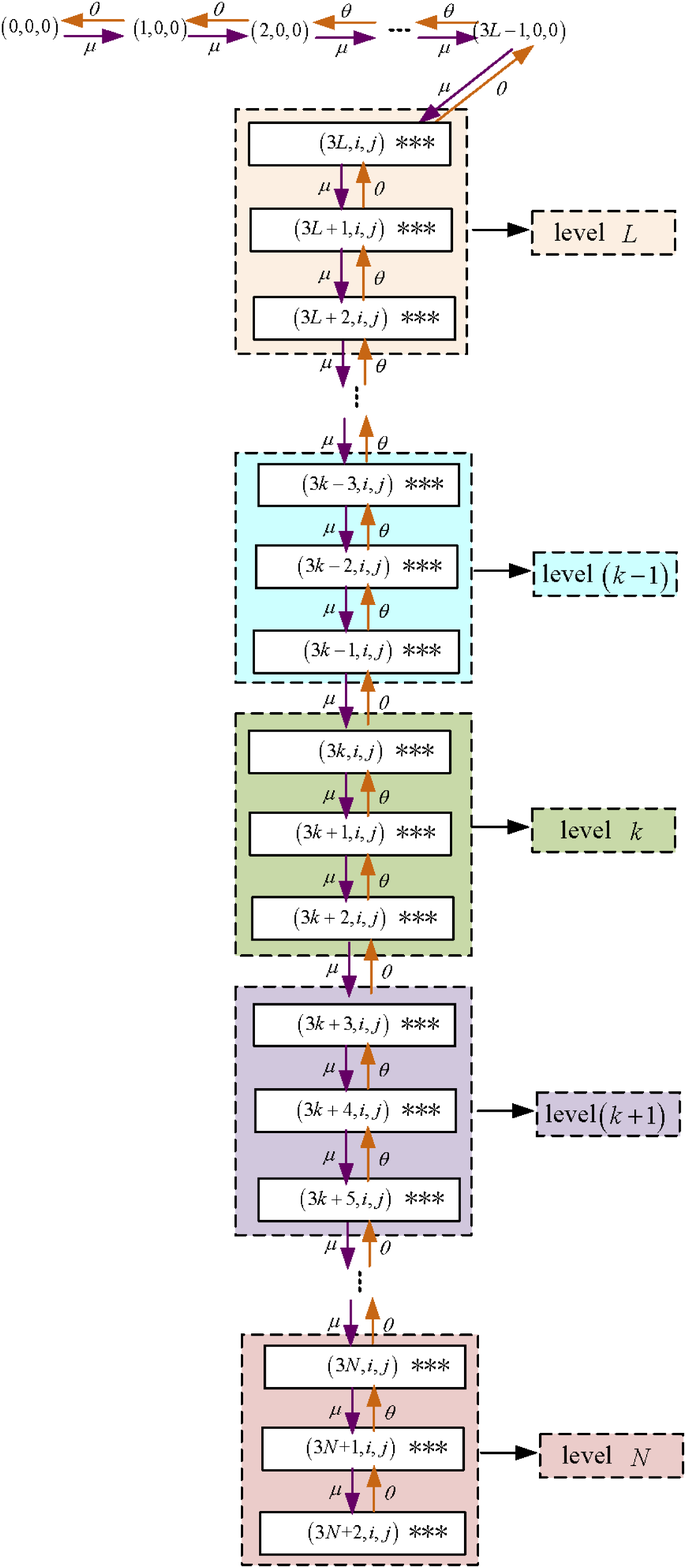}  \caption{The state
transition relations between any two levels.}%
\label{figure:Fig-1}%
\end{figure}

\begin{figure}[ptbh]
\centering         \includegraphics[width=12cm]{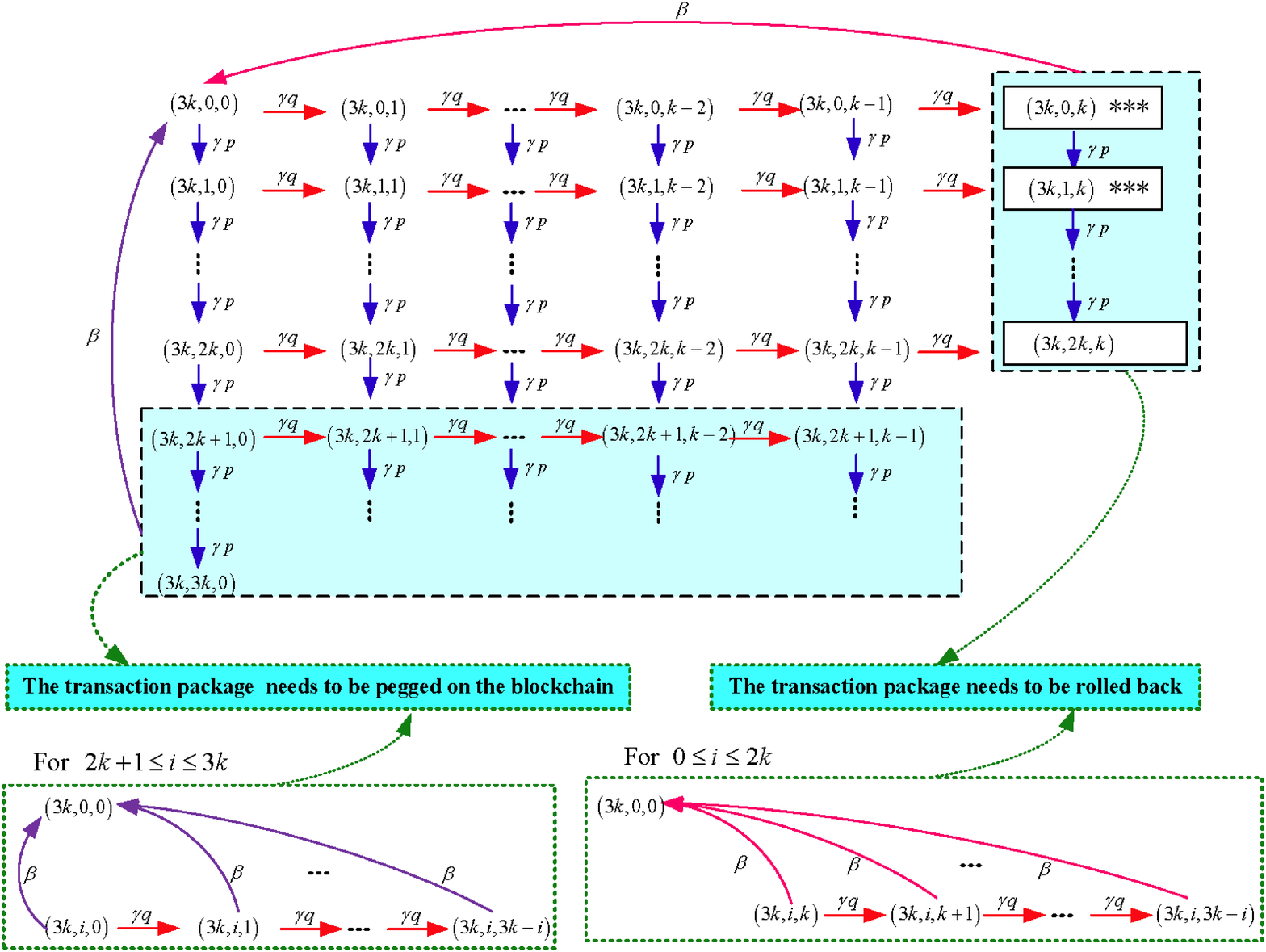}  \caption{The state
transition relations in Sublevel$_{k,0}$.}%
\label{figure:Fig-2}%
\end{figure}

\begin{figure}[ptbh]
\centering         \includegraphics[width=12cm]{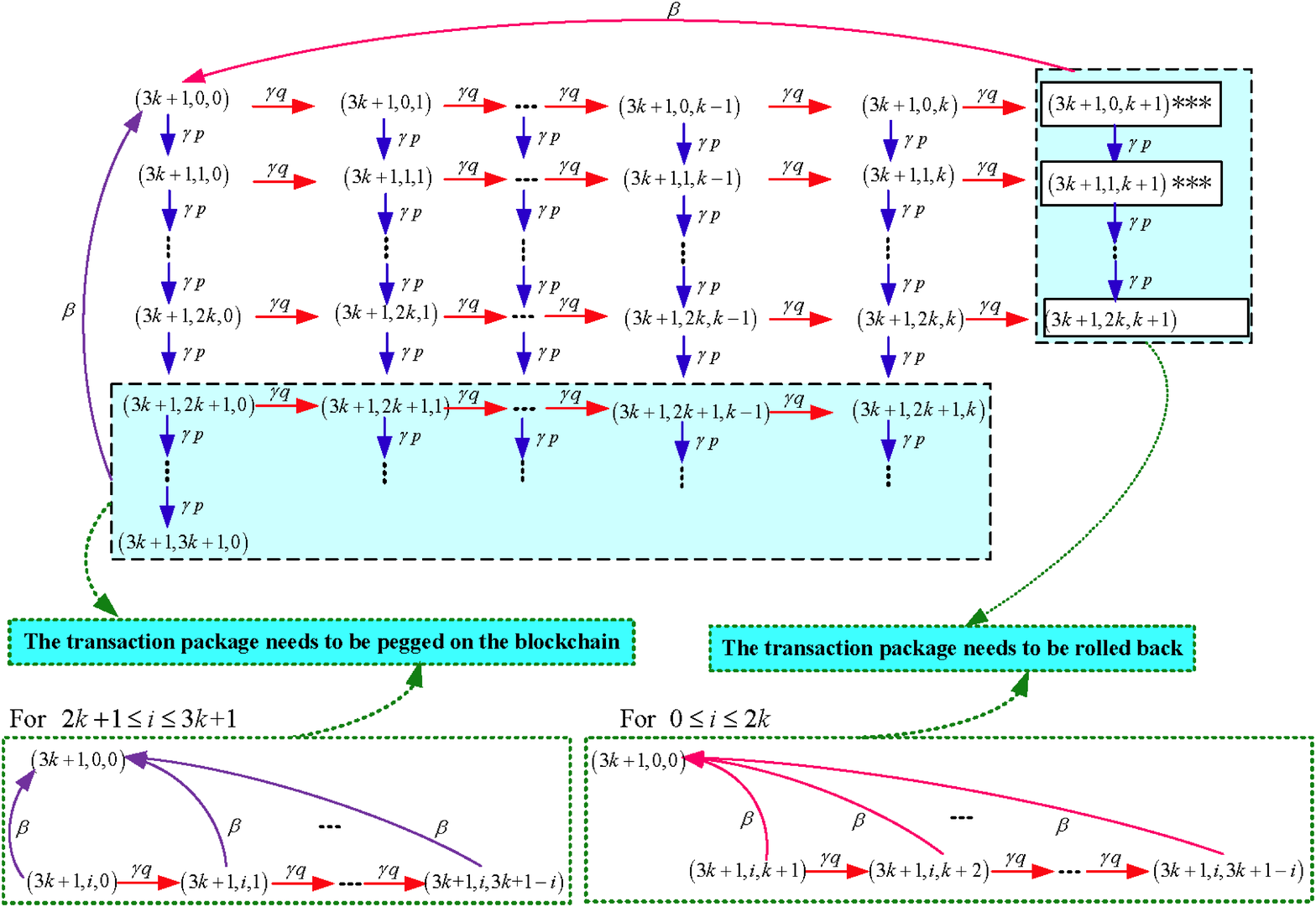}  \caption{The state
transition relations in Sublevel$_{k,1}$.}%
\label{figure:Fig-3}%
\end{figure}

\begin{figure}[ptbhptbh]
\centering         \includegraphics[width=12cm]{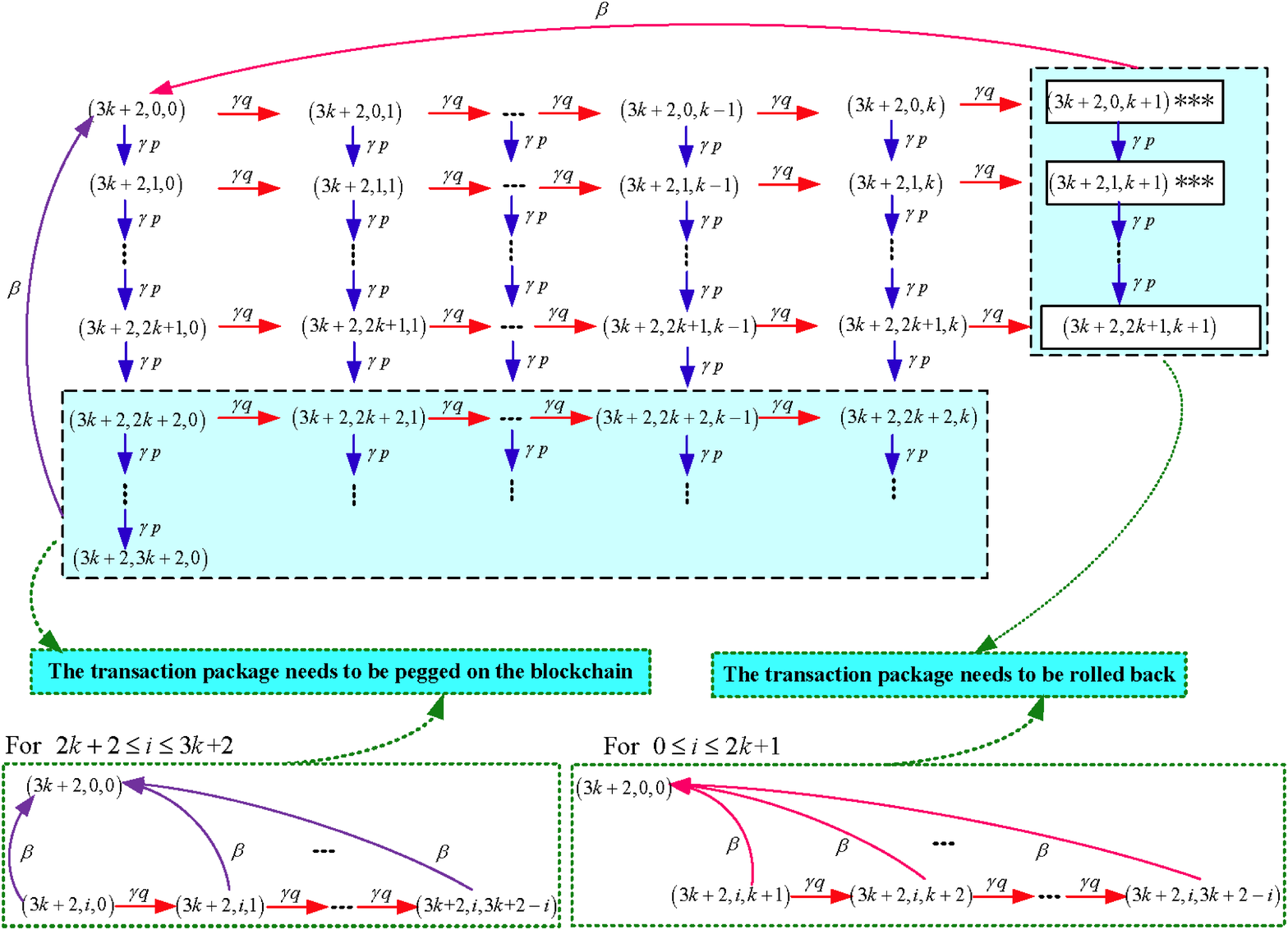}  \caption{The state
transition relations in Sublevel$_{k,2}$.}%
\label{figure:Fig-4}%
\end{figure}

The state transition relations between any two levels are depicted in Figure
\ref{figure:Fig-1}, and the state transitions in each sub-level are depicted
in Figures \ref{figure:Fig-2} to \ref{figure:Fig-4}. Note that the complicated
structure of the state transition is due to the fact that some nodes can enter
and leave the PBFT network.

From Figure \ref{figure:Fig-1} to \ref{figure:Fig-4}, it is easy to see that
\[
N(t)\in\left\{  0,1,2,\ldots,3L-1;3L,3L+1,3L+2;\ldots;3N,3N+1,3N+2\right\}  ,
\]%
\[
M(t)\in\left\{  0,1,2,\ldots,N(t)\right\}  ,
\]
\[
K(t)\in\left\{  0,1,2,\ldots,N(t)\right\}  .
\]



By using Figures \ref{figure:Fig-1} to \ref{figure:Fig-4}, the infinitesimal
generator of the Markov process $\left\{  \left(  N(t),M(t),K(t)\right)
:t\geq0\right\}  $ is given by
\begin{equation}
Q=\left(
\begin{array}
[c]{ccccccc}%
A_{1}^{(0)} & A_{0}^{(0)} &  &  &  &  & \\
A_{2}^{(L)} & A_{1}^{(L)} & A_{0}^{(L)} &  &  &  & \\
& A_{2}^{(L+1)} & A_{1}^{(L+1)} & A_{0}^{(L+1)} &  &  & \\
&  & A_{2}^{(L+2)} & A_{1}^{(L+2)} & A_{0}^{(L+2)} &  & \\
&  &  & \ddots & \ddots & \ddots & \\
&  &  &  &  &  & \\
&  &  &  & A_{2}^{(N-1)} & A_{1}^{(N-1)} & A_{0}^{(N-1)}\\
&  &  &  &  & A_{2}^{(N)} & A_{1}^{(N)}%
\end{array}
\right)  ,
\end{equation}
where
\[
A_{1}^{(0)}=\left(
\begin{array}
[c]{cccccc}%
-\mu & \mu &  &  &  & \\
\theta & -(\mu+\theta) & \mu &  &  & \\
& \theta & -(\mu+\theta) & \mu &  & \\
&  & \ddots & \ddots & \ddots & \\
&  &  & \theta & -(\mu+\theta) & \mu\\
&  &  &  & \theta & -(\mu+\theta)
\end{array}
\right)  ,
\]%
\[
A_{0}^{(0)}=\left(  Q_{0,3L},0,0\right)  ,\text{ \ }A_{2}^{(L)}=\left(
Q_{3L,0},0,0\right)  ,
\]%
\[
A_{1}^{(k)}=\left(
\begin{array}
[c]{ccc}%
Q_{3k,3k} & Q_{3k,3k+1} & \\
Q_{3k+1,3k} & Q_{3k+1,3k+1} & Q_{3k+1,3k+2}\\
& Q_{3k+2,3k+1} & Q_{3k+2,3k+2}%
\end{array}
\right)  ,\text{ \ }L\leq k\leq N,
\]%
\[
A_{2}^{(k)}=\left(  Q_{3k,3k-1},0,0\right)  ,L+1\leq k\leq N,
\]
\[
A_{0}^{(k)}=\left(  Q_{3k+2,3k+3},0,0\right)  ,\text{ }L\leq k\leq N-1.
\]

Further, the infinitesimal generator of the Markov process $\left\{  \left(
N(t),M(t),K(t)\right)  :t\geq0\right\}  $ can be rewritten as
\begin{equation}
Q=\left(
\begin{array}
[c]{ccccccccc}%
Q_{1}^{(0)} & Q_{0}^{(0)} &  &  &  &  &  &  & \\
Q_{2}^{(3L)} & Q_{1}^{(3L)} & Q_{0}^{(3L)} &  &  &  &  &  & \\
& Q_{2}^{(3L+1)} & Q_{1}^{(3L+1)} & Q_{0}^{(3L+1)} &  &  &  &  & \\
&  & Q_{2}^{(3L+2)} & Q_{1}^{(3L+2)} & Q_{0}^{(3L+2)} &  &  &  & \\
&  &  & \ddots & \ddots & \ddots &  &  & \\
&  &  &  &  & Q_{2}^{(3N+1)} & Q_{1}^{(3N+1)} & Q_{0}^{(3N+1)} & \\
&  &  &  &  &  & Q_{2}^{(3N+2)} & Q_{1}^{(3N+2)} &
\end{array}
\right)  .
\end{equation}
where
\[
Q_{1}^{(0)}=A_{1}^{(0)},\text{ \ }Q_{0}^{\left(  0\right)  }=\left(
{{{\tilde{Q}}_{0,0}},0,\ldots,0}\right)  ,
\]
\[
{\tilde{Q}_{0,0}}={\left(  {%
\begin{array}
[c]{cccc}%
0 &  &  & \\
\vdots &  &  & \\
0 &  &  & \\
\mu & 0 & \cdots & 0
\end{array}
}\right)  _{(3L)\times(3L+1)}};
\]
for $3L\leq l\leq3N+1$,
\[
Q_{0}^{(l)}=\left(
\begin{array}
[c]{ccccc}%
A_{0,0} &  &  &  & \\
& A_{1,1} &  &  & \\
&  & \ddots &  & \\
&  &  & A_{l,l} & 0
\end{array}
\right)  ,
\]%
\[
A_{i,i}=\left(
\begin{array}
[c]{ccccc}%
\mu &  &  &  & \\
& \mu &  &  & \\
&  & \ddots &  & \\
&  &  & \mu & 0
\end{array}
\right)  _{(l+1-i)\times(l+2-i)},\text{ \ \ }0\leq i\leq l;
\]
for $3L+1\leq l\leq3N+2$,%
\[
Q_{2}^{\left(  l\right)  }=\left(
\begin{array}
[c]{cccc}%
B_{0,0} &  &  & \\
& B_{1,1} &  & \\
&  & \ddots & \\
&  &  & B_{l-1,l-1}\\
&  &  & 0
\end{array}
\right)  ,
\]%
\[
B_{j,j}=\left(  {%
\begin{array}
[c]{cccc}%
\theta &  &  & \\
& \theta &  & \\
&  & \ddots & \\
&  &  & \theta\\
&  &  & 0
\end{array}
}\right)  _{(l+1-j)\times(l-j)},\text{ \ }0\leq j\leq l-1;
\]%
\[
Q_{2}^{\left(  3L\right)  }=\left(  {%
\begin{array}
[c]{c}%
\tilde{\tilde{Q}}_{0,0}\\
0\\
\vdots\\
0
\end{array}
}\right)  ,
\]%
\[
\tilde{\tilde{Q}}_{0,0}=\left(  {%
\begin{array}
[c]{cccc}%
0 & \cdots & 0 & \theta\\
0 &  &  & \\
\vdots &  &  & \\
0 &  &  &
\end{array}
}\right)  _{(3L+1)\times(3L)};
\]
for $3L\leq l\leq3N+2$,%
\[
Q_{1}^{\left(  l\right)  }=\left(  {%
\begin{array}
[c]{cccccc}%
C_{0,0} & C_{0,1} &  &  &  & \\
C_{1,0} & C_{1,1} & C_{1,2} &  &  & \\
C_{2,0} &  & C_{2,2} & C_{2,3} &  & \\
\vdots &  &  & \ddots & \ddots & \\
C_{l-1,0} &  &  &  & C_{l-1,l-1} & C_{l-1,l}\\
C_{l,0} &  &  &  &  & C_{l,l}%
\end{array}
}\right)  ,
\]
for $3L\leq l\leq3N+1$,
\[
{C_{l,l}}=-(\mu+\beta),\text{ \ }{C_{l,0}}=\left(  {\beta,0,0,\ldots
,0}\right)  ,
\]
for $l=3N+2$,
\[
{C_{l,l}}=-\beta,\quad{C_{l,0}}=\left(  {\beta,0,0,\ldots,0}\right)  ;
\]
for $3L\leq l\leq3N+2$ and $r=0,1,2,\ldots,l-1$,%
\[
C_{r,r+1}=\left(  {%
\begin{array}
[c]{cccc}%
\gamma p &  &  & \\
& \gamma p &  & \\
&  & \ddots & \\
&  &  & \gamma p\\
&  &  & 0
\end{array}
}\right)  _{(l+1-r)\times(l-r)},
\]%
\[
C_{0,0}=\left(  {%
\begin{array}
[c]{c}%
D_{0,0}\\
E_{0,0}%
\end{array}
}\right)  ,E_{0,0}=\left(  {%
\begin{array}
[c]{cc}%
{{F_{0,0}}} & {{G_{0,0}}}%
\end{array}
}\right)  ,F_{0,0}=\left(  {%
\begin{array}
[c]{cc}%
\beta & \\
\vdots & \\
\beta &
\end{array}
}\right)  _{(l+1-{k_{l}})\times{k_{l}}},
\]%
\[
G_{0,0}=\left(  {%
\begin{array}
[c]{cccc}%
{-(\gamma+\mu+\theta+\beta)} & {\gamma q} &  & \\
& \ddots & \ddots & \\
&  & {-(\gamma+\mu+\theta+\beta)} & {\gamma q}\\
&  &  & {-(\mu+\beta)}%
\end{array}
}\right)  _{(l+1-{k_{l}})\times(l+1-{k_{l}})},
\]
\[
D_{0,0}=\left(  {%
\begin{array}
[c]{cccccc}%
{-(\gamma+\theta+\mu)} & {\gamma q} &  &  &  & \\
& {-(\gamma+\theta+\mu)} & {\gamma q} &  &  & \\
&  & \ddots & \ddots &  & \\
&  &  & {-(\gamma+\theta+\mu)} & {\gamma q} &
\end{array}
}\right)  _{{k_{l}}\times(l+1)},
\]
for $l=3L$,
\[
G_{0,0}=\left(  {%
\begin{array}
[c]{cccc}%
{-(\gamma+\mu+\beta)} & {\gamma q} &  & \\
& \ddots & \ddots & \\
&  & {-(\gamma+\mu+\beta)} & {\gamma q}\\
&  &  & {-(\mu+\beta)}%
\end{array}
}\right)  _{(2L+1)\times(2L+1)},
\]
\[
D_{0,0}=\left(  {%
\begin{array}
[c]{cccccc}%
{-(\gamma+\mu+\theta)} & {\gamma q} &  &  &  & \\
& {-(\gamma+\mu)} & {\gamma q} &  &  & \\
&  & \ddots & \ddots &  & \\
&  &  & {-(\gamma+\mu)} & {\gamma q} &
\end{array}
}\right)  _{L\times(3L+1)},
\]
for $l=3N+2$,
\[
G_{0,0}=\left(  {%
\begin{array}
[c]{cccc}%
{-(\gamma+\theta+\beta)} & {\gamma q} &  & \\
& \ddots & \ddots & \\
&  & {-(\gamma+\theta+\beta)} & {\gamma q}\\
&  &  & {-\beta}%
\end{array}
}\right)  _{(2N+2)\times(2N+2)};
\]
\[
D_{0,0}=\left(  {%
\begin{array}
[c]{cccccc}%
{-(\gamma+\theta)} & {\gamma q} &  &  &  & \\
& {-(\gamma+\theta)} & {\gamma q} &  &  & \\
&  & \ddots & \ddots &  & \\
&  &  & {-(\gamma+\theta)} & {\gamma q} &
\end{array}
}\right)  _{(N+1)\times(3N+3)},
\]
for $r=1,2,\ldots,m_{l}-2$,
\[
C_{r,0}=\left(  {%
\begin{array}
[c]{c}%
0\\
{{H_{r,0}}}%
\end{array}
}\right)  _{(l+1-r)\times(l+1)},H_{r,0}=\left(  {%
\begin{array}
[c]{cc}%
\beta & \\
\beta & \\
\vdots & \\
\beta &
\end{array}
}\right)  _{(l+1-{k_{l}}-r)\times(l+1)},
\]%
\[
C_{r,r}=\left(  {%
\begin{array}
[c]{c}%
{{I_{r,r}}}\\
{{J_{r,r}}}%
\end{array}
}\right)  ,J_{r,r}=\left(  {%
\begin{array}
[c]{cc}%
0 & {{K_{r,r}}}%
\end{array}
}\right)  ,
\]%
\[
I_{r,r}=\left(  {%
\begin{array}
[c]{cccccc}%
{-(\gamma+\theta+\mu)} & {\gamma q} &  &  &  & \\
& {-(\gamma+\theta+\mu)} & {\gamma q} &  &  & \\
&  & \ddots & \ddots &  & \\
&  &  & {-(\gamma+\theta+\mu)} & {\gamma q} &
\end{array}
}\right)  _{{k_{l}}\times(l+1-r)},
\]%
\[
K_{r,r}=\left(  {%
\begin{array}
[c]{cccc}%
{-(\gamma+\mu+\theta+\beta)} & {\gamma q} &  & \\
& \ddots & \ddots & \\
&  & {-(\gamma+\mu+\theta+\beta)} & {\gamma q}\\
&  &  & {-(\mu+\beta)}%
\end{array}
}\right)  _{(l+1-{k_{l}}-r)\times(l+1-{k_{l}}-r)},
\]
for $l=3L$,
\[
I_{r,r}=\left(  {%
\begin{array}
[c]{cccccc}%
{-(\gamma+\mu)} & {\gamma q} &  &  &  & \\
& {-(\gamma+\mu)} & {\gamma q} &  &  & \\
&  & \ddots & \ddots &  & \\
&  &  & {-(\gamma+\mu)} & {\gamma q} &
\end{array}
}\right)  _{L\times(3L+1-r)},
\]%
\[
K_{r,r}=\left(  {%
\begin{array}
[c]{cccc}%
{-(\gamma+\mu+\beta)} & {\gamma q} &  & \\
& \ddots & \ddots & \\
&  & {-(\gamma+\mu+\beta)} & {\gamma q}\\
&  &  & {-(\mu+\beta)}%
\end{array}
}\right)  _{(2L+1-r)\times(2L+1-r)},
\]
for $l=3N+2$,
\[
I_{r,r}=\left(  {%
\begin{array}
[c]{cccccc}%
{-(\gamma+\theta)} & {\gamma q} &  &  &  & \\
& {-(\gamma+\theta)} & {\gamma q} &  &  & \\
&  & \ddots & \ddots &  & \\
&  &  & {-(\gamma+\theta)} & {\gamma q} &
\end{array}
}\right)  _{(N+1)\times(3N+3-r)},
\]%
\[
K_{r,r}=\left(  {%
\begin{array}
[c]{cccc}%
{-(\gamma+\theta+\beta)} & {\gamma q} &  & \\
& \ddots & \ddots & \\
&  & {-(\gamma+\theta+\beta)} & {\gamma q}\\
&  &  & {-\beta}%
\end{array}
}\right)  _{(2N+2-r)\times(2N+2-r)};
\]
for $r={m_{l}}-1$,
\[
C_{r,0}=\left(  {%
\begin{array}
[c]{cc}
& \\
\beta &
\end{array}
}\right)  _{(l+1-r)\times(l+1)},
\]
\[
C_{r,r}=\left(  {%
\begin{array}
[c]{cccc}%
{-(\gamma+\mu+\theta)} & {\gamma q} &  & \\
& \ddots & \ddots & \\
&  & {-(\gamma+\mu+\theta)} & {\gamma q}\\
&  &  & {-(\mu+\beta)}%
\end{array}
}\right)  _{(l+1-r)\times(l+1-r)},
\]
for $l=3L$,
\[
C_{r,r}=\left(  {%
\begin{array}
[c]{ccccc}%
{-(\gamma+\mu)} & {\gamma q} &  &  & \\
& {-(\gamma+\mu)} & {\gamma q} &  & \\
&  & \ddots & \ddots & \\
&  &  & {-(\gamma+\mu)} & {\gamma q}\\
&  &  &  & {-(\mu+\beta)}%
\end{array}
}\right)  _{(3L+1-r)\times(3L+1-r)},
\]
for $l=3N+2$,
\[
C_{r,r}=\left(  {%
\begin{array}
[c]{ccccc}%
{-(\gamma+\theta)} & {\gamma q} &  &  & \\
& \ddots & \ddots &  & \\
&  & {-(\gamma+\theta)} & {\gamma q} & \\
&  &  & {-(\gamma+\theta)} & {\gamma q}\\
&  &  &  & {-\beta}%
\end{array}
}\right)  _{(l+1-r)\times(l+1-r)};
\]
for $m_{l}\leq r\leq l-1$,
\[
C_{r,0}=\left(  {%
\begin{array}
[c]{cc}%
\beta & \\
\vdots & \\
\beta &
\end{array}
}\right)  _{(l+1-r)\times(l+1)},
\]%
\[
C_{r,r}=\left(  {%
\begin{array}
[c]{cccc}%
{-(\gamma+\mu+\theta+\beta)} & {\gamma q} &  & \\
& \ddots & \ddots & \\
&  & {-(\gamma+\mu+\theta+\beta)} & {\gamma q}\\
&  &  & {-(\mu+\beta)}%
\end{array}
}\right)  _{(l+1-r)\times(l+1-r)},
\]
for $l=3L$,
\[
C_{r,r}=\left(  {%
\begin{array}
[c]{ccccc}%
{-(\gamma+\mu+\beta)} & {\gamma q} &  &  & \\
& {-(\gamma+\mu+\beta)} & {\gamma q} &  & \\
&  & \ddots & \ddots & \\
&  &  & {-(\gamma+\mu+\beta)} & {\gamma q}\\
&  &  &  & {-(\mu+\beta)}%
\end{array}
}\right)  _{(l+1-r)\times(l+1-r)},
\]
for $l=3N+2$,
\[
C_{r,r}=\left(  {%
\begin{array}
[c]{ccccc}%
{-(\gamma+\theta+\beta)} & {\gamma q} &  &  & \\
& {-(\gamma+\theta+\beta)} & {\gamma q} &  & \\
&  & \ddots & \ddots & \\
&  &  & {-(\gamma+\theta+\beta)} & {\gamma q}\\
&  &  &  & {-\beta}%
\end{array}
}\right)  _{(l+1-r)\times(l+1-r)}.
\]

\begin{Rem}
Although the Markov process of the dynamic PBFT voting process is more
complicated, we can still write the state transition relations, and the
infinitesimal generator of the QBD process. This is a key step in our
subsequent study, for example, performance analysis, and numerical computation.
\end{Rem}

\section{Performance Analysis for the Dynamic PBFT Voting Process}

\label{sec:performnace measures}

In this section, we first provide the stationary probability vector of the QBD
process. Then we provide performance analysis for the dynamic PBFT voting process.

\subsection{The stationary probability vector}

Note that the QBD process $Q$ is irreducible and contains finite states, thus
it is positive recurrent. Let ${\boldsymbol{\pi}  }$ be the stationary
probability vector of the QBD process $Q$. Based on Figures \ref{figure:Fig-1}
to \ref{figure:Fig-4}, we write%
\[
{\boldsymbol{\pi}  }=\left(  {{\pi_{0}},{\pi_{3L}},{\pi_{3L+1}},{\pi_{3L+2}%
},\ldots,{\pi_{3N}},{\pi_{3N+1}},{\pi_{3N+2}}}\right)  ,
\]
where
\[
{{\boldsymbol{\pi}  }_{0}}=\left(  {{\pi_{0,0,0}},{\pi_{1,0,0}},{\pi_{2,0,0}%
},\ldots,{\pi_{3L-1,0,0}}}\right)  ,
\]%
\[
{{\boldsymbol{\pi}  }_{3L}}=\left(  {{\pi_{3L,0,0}},{\pi_{3L,0,1}},\ldots
,{\pi_{3L,0,3L}};{\pi_{3L,1,0}},{\pi_{3L,1,1}},\ldots,{\pi_{3L,1,3L-1}}%
;\ldots;{\pi_{3L,3L,0}}}\right)  ,
\]%
\[
{{\boldsymbol{\pi}  }_{3L+1}}=\left(  {{\pi_{3L+1,0,0}},{\pi_{3L+1,0,1}%
},\ldots,{\pi_{3L+1,0,3L+1}};{\pi_{3L+1,1,0}},\ldots,{\pi_{3L+1,1,3L}}%
;\ldots;{\pi_{3L+1,3L+1,0}}}\right)  ,
\]%
\[
\vdots
\]%
\[
{{\boldsymbol{\pi}  }_{3N+1}}=\left(  {{\pi_{3N+1,0,0}},{\pi_{3N+1,0,1}%
},\ldots,{\pi_{3N+1,0,3N+1}};{\pi_{3N+1,1,0}},\ldots,{\pi_{3N+1,1,3N}}%
;\ldots;{\pi_{3N+1,3N+1,0}}}\right)  ,
\]%
\[
{{\boldsymbol{\pi}  }_{3N+2}}=\left(  {{\pi_{3N+2,0,0}},{\pi_{3N+2,0,1}%
},\ldots,{\pi_{3N+2,0,3N+2}};{\pi_{3N+2,1,0}},\ldots,{\pi_{3N+2,1,3N+1}%
};\ldots;{\pi_{3N+2,3N+2,0}}}\right)  .
\]

Note that the stationary probability vector $\boldsymbol{\pi}  $ can be
obtained by means of solving the system of linear equations $\boldsymbol{\pi}
Q={\mathbf{0}}$ and $\boldsymbol{\pi}  \mathbf{e}=1$, where $\mathbf{e}$ is a
column vector of the ones with a suitable size.

Now, we use the UL-type RG-factorization to compute the stationary probability
vector $\boldsymbol{\pi}  $ as follows.

We write
\begin{equation}
{U_{3N + 2}} = Q_{1}^{(3N + 2)}, \label{e6}%
\end{equation}
\begin{equation}
{U_{k}} = Q_{1}^{(k)} + Q_{0}^{(k)}\left(  { -U_{k + 1}^{ -1}} \right)
Q_{2}^{(k + 1)}, \quad3L \le k \le3N + 1, \label{e7}%
\end{equation}
\begin{equation}
{U_{0}} = Q_{1}^{(0)} + Q_{0}^{(0)}\left(  { -U_{3L}^{ -1}} \right)
Q_{2}^{(3L)}. \label{e8}%
\end{equation}
Based on the U-measure $\left\{  {{U_{k}}} \right\}  $, we can respectively
define the UL-type R- and G-measures as
\begin{equation}
{R_{0}} = Q_{0}^{(0)}\left(  { -U_{3L}^{ -1}} \right)  , \label{e9}%
\end{equation}
\begin{equation}
{R_{k}} = Q_{0}^{(k)}\left(  { -U_{k + 1}^{ -1}} \right)  , \quad3L \le k
\le3N + 1, \label{e10}%
\end{equation}
and
\begin{equation}
{G_{k}} = \left(  { - U_{k}^{ - 1}} \right)  Q_{2}^{(k)},\quad3L \le k \le3N +
2. \label{e11}%
\end{equation}

Note that the matrix sequence $\left\{  R_{0},R_{3L},R_{3L+1},\ldots
,R_{3N}\right\}  $ is the unique nonnegative solution to the system of matrix
equations
\[
\left\{
\begin{array}
[c]{l}%
Q_{0}^{(0)}+R_{0}Q_{1}^{(3L)}+R_{0}R_{3L}Q_{2}^{(3L+1)}=0,\\
Q_{0}^{(k)}+R_{k}Q_{1}^{(k+1)}+R_{k}R_{k+1}Q_{2}^{(k+2)}=0,\quad3L\leq
k\leq3N,
\end{array}
\right.
\]
with the boundary condition
\[
R_{3N+1}=Q_{0}^{(3N+1)}\left(  -U_{3N+2}^{-1}\right)  .
\]
Hence we obtain
\[
\left\{
\begin{array}
[c]{l}%
R_{0}=-Q_{0}^{(0)}\left[  Q_{1}^{(3L)}+R_{3L}Q_{2}^{(3L+1)}\right]  ^{-1},\\
R_{k}=-Q_{0}^{(k)}\left[  Q_{1}^{(k+1)}+R_{k+1}Q_{2}^{(k+2)}\right]
^{-1},\quad3L\leq k\leq3N.
\end{array}
\right.
\]
Similarly, the matrix sequence $\left\{  G_{k},3L\leq k\leq3N+1\right\}  $ is
the unique nonnegative solution to the system of matrix equations
\[
Q_{0}^{(k)}G_{k+1}G_{k}+Q_{1}^{(k)}G_{k}+Q_{2}^{(k)}=0,\quad3L\leq k\leq3N+1,
\]
with the boundary condition
\[
G_{3N+2}=\left(  -U_{3N+2}^{-1}\right)  Q_{2}^{(3N+2)}.
\]
Thus
\[
G_{k}=-\left[  Q_{0}^{(k)}G_{k+1}+Q_{1}^{(k)}\right]  ^{-1} Q_{2}^{(k)}%
,\quad3L\leq k\leq3N+1.
\]

For the QBD process $Q$ with finitely-many levels, the UL-type
RG-factorization is given by
\[
Q=\left(  I-R_{U}\right)  U_{D}\left(  I-G_{L}\right)  ,
\]
where
\[
R_{U}=\left(  {%
\begin{array}
[c]{cccccc}%
0 & R_{0} &  &  &  & \\
& 0 & R_{\mathrm{{3}}L} &  &  & \\
&  & \ddots & \ddots &  & \\
&  &  & 0 & R_{\mathrm{{3}}N} & \\
&  &  &  & 0 & R_{\mathrm{{3}}N\mathrm{{+}}1}\\
&  &  &  &  & 0
\end{array}
}\right)  ,
\]%
\[
U_{D}=\text{diag}\left(  {{U_{0}},{U_{\mathrm{{3}}L}},{U_{\mathrm{{3}}L+1}%
},\ldots,{U_{\mathrm{{3}}N\mathrm{{+}}1}},{U_{\mathrm{{3}}N\mathrm{{+2}}}}%
}\right)  ,
\]%
\[
G_{L}=\left(  {%
\begin{array}
[c]{cccccc}%
0 &  &  &  &  & \\
{{G_{\mathrm{{3}}L}}} & 0 &  &  &  & \\
& {{G_{\mathrm{{3}}L+1}}} & 0 &  &  & \\
&  & \ddots & \ddots &  & \\
&  &  & {{G_{\mathrm{{3}}N\mathrm{{+}}1}}} & 0 & \\
&  &  &  & {{G_{\mathrm{{3}}N\mathrm{{+2}}}}} & 0
\end{array}
}\right)  .
\]

Using the Chapter 2 in Li \cite{Li:2010}, the following theorem provides the
stationary probability vector of the Markov process $Q$, and its proof is easy
and is omitted here.

\begin{The}
\label{the1} The stationary probability vector of the Markov process $Q$ is
given by
\begin{equation}
\left\{
\begin{array}
[c]{l}%
\boldsymbol{\pi}  _{0}=\varphi v_{0},\\
\boldsymbol{\pi}  _{3 L}=\varphi v_{0} R_{0},\\
\boldsymbol{\pi}  _{k}=\varphi v_{0} R_{0} R_{1} \cdots R_{k-1}, \quad3L+1
\leq k \leq3 N+2 . \label{e12}%
\end{array}
\right.
\end{equation}
where, $v_{0}$ is the stationary probability vector of the censored Markov
chain $U_{0}=Q_{1}^{(0)}+R_{0}Q_{2}^{(3L)}$ to level $0$, and the positive
scalar $\varphi$ is regularization constant and it is uniquely determined by
\[
{{\boldsymbol{\pi}  }_{0}}\mathbf{{e}}+\sum\limits_{k=3L}^{3N+2}%
{{{\boldsymbol{\pi}  }_{k}}{\mathbf{e}}}=1.
\]
\end{The}

\subsection{Performance Analysis}

Using the stationary probability vector $\boldsymbol{\pi} $ given in Theorem
\ref{the1}, we can provide some performance measures of the dynamic PBFT
voting process as follows:

\textbf{(a) The stationary probability that the transaction package becomes a
block is given by}
\[
\zeta_{1}=\sum\limits_{l=3L}^{3N+2}{\sum\limits_{m\geq{m_{l}}}^{l}%
{\sum\limits_{k=0}^{l-m}{{\pi_{l,m,k}}}}}.
\]

\textbf{(b) The stationary probability that the transaction package becomes an
orphan block is given by}
\[
\zeta_{2}=\sum\limits_{l=3L}^{3N+2}{\sum\limits_{m=0}^{{m_{l}}-1}%
{\sum\limits_{k\geq{k_{l}}}^{l-m}{{\pi_{l,m,k}}}}}.
\]

\textbf{(c) (i) The stationary probability that the dynamic PBFT system
completes the voting process is given by}
\[
\mathbf{{A}}=\sum\limits_{l=3L}^{3N+2}{\sum\limits_{m\geq{m_{l}}}^{l}%
{\sum\limits_{k=0}^{l-m}{{\pi_{l,m,k}}}}}+\sum\limits_{l=3L}^{3N+2}%
{\sum\limits_{m=0}^{{m_{l}}-1}{\sum\limits_{k\geq{k_{l}}}^{l-m}{{\pi_{l,m,k}}%
}}}={\zeta_{1}}+{\zeta_{2}}.
\]

\textbf{(ii) The stationary probability that the dynamic PBFT system cannot
perform the voting process is given by}
\[
\mathbf{{B}}=\sum\limits_{i=0}^{3L-1}{{\pi_{i,0,0}}}=\boldsymbol{\pi}
_{0}\mathbf{e}.
\]

\textbf{(iii) The stationary probability that the dynamic PBFT system perform
the voting process but it cannot complete the voting process is given by}%
\[
\mathbf{{C}}=1-\mathbf{{A}}-\mathbf{{B}}.
\]

\textbf{(d) The stationary rate that the blocks are pegged on the blockchain
is given by}
\begin{equation}
{r_{1}}\mathrm{{ = }}\beta\left(  {\sum\limits_{l = 3L}^{3N + 2}
{\sum\limits_{m \ge{m_{l}}}^{l} {\sum\limits_{k = 0}^{l - m} {{\pi_{l,m,k}}} }
} } \right)  = \beta{\zeta_{1}}. \label{e13}%
\end{equation}

\textbf{(e) The stationary rate the the orphan blocks are rolled back is given
by}
\begin{equation}
{r_{2}}\mathrm{{ = }}\beta\left(  {\sum\limits_{l = 3L}^{3N + 2}
{\sum\limits_{m = 0}^{{m_{l}} - 1} {\sum\limits_{k \ge{k_{l}}}^{l - m}
{{\pi_{l,m,k}}} } } } \right)  = \beta{\zeta_{2}}. \label{e14}%
\end{equation}

\section{The dynamic PBFT Blockchain System}

\label{sec:dynamic system}

In this section, we set up an $\mathrm{M}\oplus\mathrm{M}^{\mathrm{b}%
}/\mathrm{M}^{\mathrm{b}}/1$ queue to approximately study the dynamic PBFT
blockchain system. Using such an approximate queueing model, we can provide
performance analysis of the dynamic PBFT blockchain system, for example, the
throughput, and the growth rate of blockchain.

\subsection{An approximate queueing model}

Note that the dynamic PBFT blockchain system is a complicated network due to
the dynamic voting processes, thus its performance analysis is always more
interesting and challenging. For this reason, we design an $\mathrm{M}%
\oplus\mathrm{M}^{\mathrm{b}}/\mathrm{M}^{\mathrm{b}}/1$ queue to
approximately analyze performance measures of the dynamic PBFT blockchain
system. Now, the $\mathrm{M}\oplus\mathrm{M}^{\mathrm{b}}/\mathrm{M}%
^{\mathrm{b}}/1$ queue is described as follows:

\textbf{(1) Transaction arrivals at the transaction pool:} We assume that the
external transactions arrive at the transaction pool according to a Poisson
process with arrival rate $\lambda>0$. See Assumption (8) in Section
\ref{sec:model descrition}.

\textbf{(2) The total arrival process: }From (e) of Subsection 4.2, we can see
that the stationary rate that all the orphan blocks are returned to the
transaction pool is given by
\[
r_{2}=\beta\left(  {\sum\limits_{l=3L}^{3N+2}{\sum\limits_{m=0}^{{m_{l}}%
-1}{\sum\limits_{k\geq{k_{l}}}^{l-m}{{\pi_{l,m,k}}}}}}\right)  =\beta
{\zeta_{2}}%
\]
with a batch size $b$ of transactions. Combining the above \textbf{(1)}, we
can get that the total transaction arrivals at this system are a composite
process between the two Poisson processes: One with arrival rate $\lambda$
while the other with arrival rate $r_{2}$, as well as batch size $b$.

\textbf{(3) The service times:} Note that the dynamic PBFT blockchain system
randomly selects $b$ transactions from the transaction pool with equal
probability to make a new transaction package of batch size $b$, and then the
transaction package immediately performs through the dynamic PBFT voting
process. We assume that the approval time of the transaction package in the
voting process is exponential with approval rate $r_{1}$.

From the perspective of queueing theory, the service time is exponential with
approval rate $r_{1}$ and batch size $b$ of transactions. That is, the
block-pegged rate is $r_{1}$.

\textbf{(4) Independence: }We assume that all random variables defined above
are independent of each other.

\begin{figure}[ptbh]
\centering         \includegraphics[width=10cm]{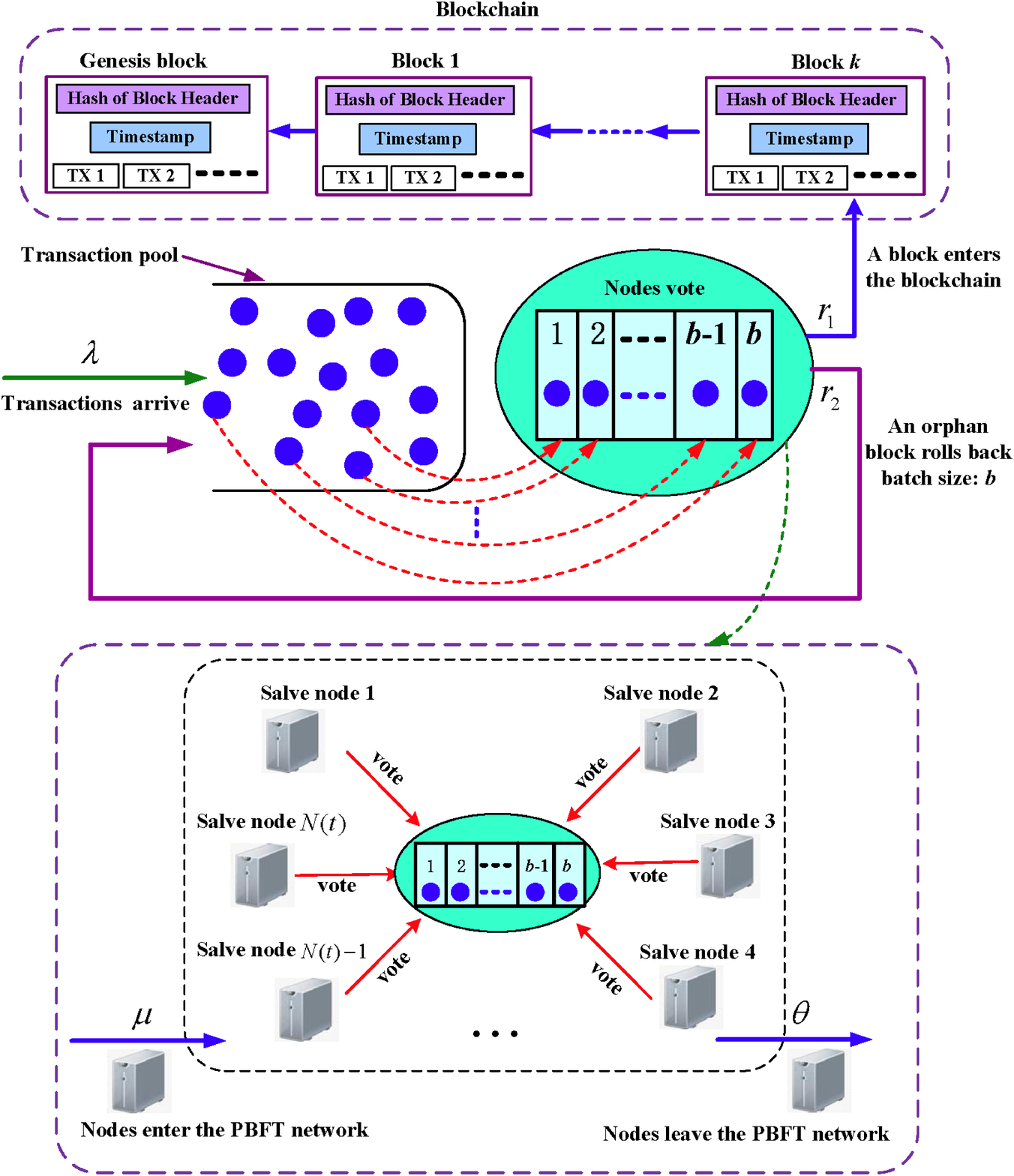}  \caption{The
$\mathrm{M}\oplus\mathrm{M}^{\mathrm{b}}/\mathrm{M}^{\mathrm{b}}/1$ queue and
its dynamic PBFT blockchain system.}%
\label{figure:Fig-5}%
\end{figure}

From the above model assumptions, it is easy to see that the dynamic PBFT
blockchain system is approximately described as an $\mathrm{M}\oplus
\mathrm{M}^{\mathrm{b}}/\mathrm{M}^{\mathrm{b}}/1$ queue. The $\mathrm{M}%
\oplus\mathrm{M}^{\mathrm{b}}/\mathrm{M}^{\mathrm{b}}/1$ queue and its dynamic
PBFT blockchain system are depicted in Figure \ref{figure:Fig-5}.

Let $I(t)$ be the number of transactions in the transaction pool at time $t$.
If $0\leq I(t)\leq b-1$, then a transaction package cannot be completed so
that the voting process will not be set up, i.e., the dynamic PBFT system is
in an idle period because there is no voting process. If $I(t)\geq b$, then
the dynamic PBFT blockchain system is in a busy period.

\begin{Rem}
Note that $r_{1}$ and $r_{2}$ are two exponentially service rates, which are
approximately obtained in the dynamic PBFT voting process according to the
transaction packages of batch size $b$.
\end{Rem}

\subsection{Analysis of the $\mathbf{M\oplus M}^{\mathbf{b}}\mathbf{/M}%
^{\mathbf{b}}\mathbf{/1}$ queue}

Now, we analyze the $\mathrm{M}\oplus\mathrm{M}^{\mathrm{b}}/\mathrm{M}%
^{\mathrm{b}}/1$ queue. It is easy to see that $\left\{  I(t):t\geq0\right\}
$ is a continuous-time Markov process whose state space is given by
\[
\Omega=\left\{  0,1,2,\ldots,b-1,b,b+1,b+2,\ldots\right\}  .
\]
Also, the state transition relations of the Markov process $\left\{
I(t):t\geq0\right\}  $ are depicted as follows. \begin{figure}[ptbh]
\centering         \includegraphics[width=12cm]{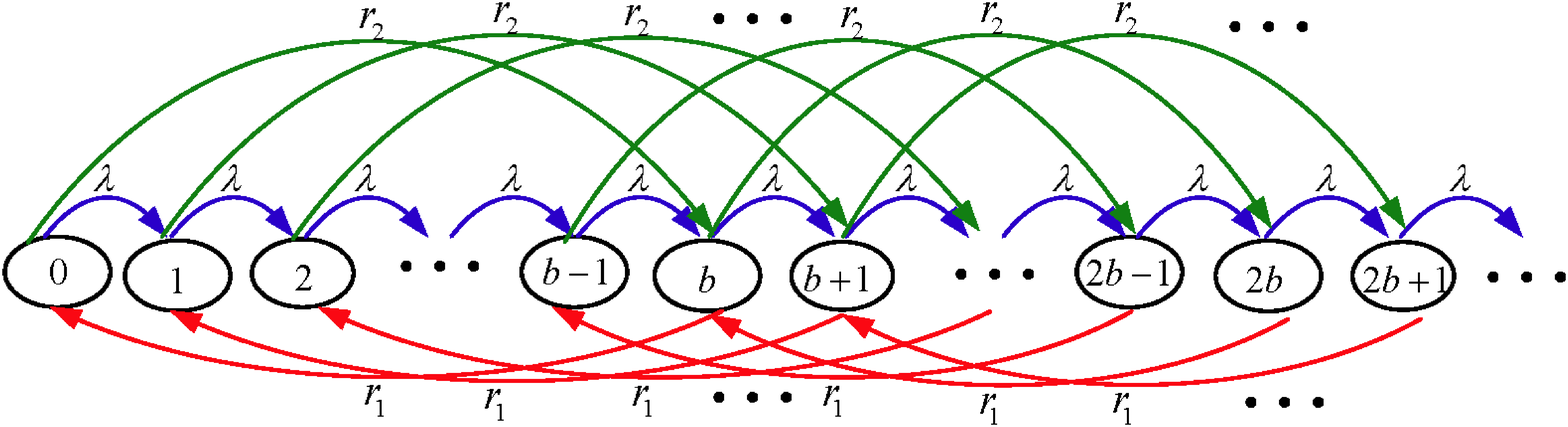}  \caption{The state
transition relations of the Markov process.}%
\label{figure:Fig-6}%
\end{figure}

Based on Figure \ref{figure:Fig-6}, the infinitesimal generator $T$ of the
Markov process $\left\{  I(t):t\geq0\right\}  $ is given by
\[
T=\left(  {%
\begin{array}
[c]{ccccccccc}%
-\left(  \lambda+r_{2}\right)   & \lambda &  & r_{2} &  &  &  &  & \\
& \ddots & \ddots &  & \ddots &  &  &  & \\
&  & {-\left(  {\lambda+{r_{2}}}\right)  } & \lambda &  & {{r_{2}}} &  &  & \\
{{r_{1}}} &  &  & {-\left(  {\lambda+{r_{2}}+{r_{1}}}\right)  } & \lambda &  &
{{r_{2}}} &  & \\
& {{r_{1}}} &  &  & {-\left(  {\lambda+{r_{2}}+{r_{1}}}\right)  } & \lambda &
& {{r_{2}}} & \\
&  & \ddots &  &  & \ddots & \ddots &  & \ddots
\end{array}
}\right)  .
\]
Let $T=\left(  T_{i,j}\right)  $. Then the elements%
\[
T_{i,i}=\left\{
\begin{array}
[c]{cc}%
-\left(  \lambda+r_{2}\right)  , & \text{if }0\leq i\leq b-1,\\
-\left(  \lambda+r_{2}+r_{1}\right)  , & \text{if }i\geq b,
\end{array}
\right.
\]%
\[
T_{i,i+1}=\lambda;\text{ \ }T_{i,i+b}=r_{2}\text{ if }i\geq0;\text{
\ }T_{i,i-b}=r_{1}\text{ if }i\geq b.
\]
Further, the infinitesimal generator $T$ can be rewritten as
\begin{equation}
\mathbf{{T}}=\left(  {%
\begin{array}
[c]{ccccc}%
{B_{1}^{(0)}} & {{A_{0}}} &  &  & \\
A_{2} & A_{1} & {{A_{0}}} &  & \\
& A_{2} & A_{1} & {{A_{0}}} & \\
&  & \ddots & \ddots & \ddots
\end{array}
}\right)  ,\label{e15}%
\end{equation}
where
\[
B_{1}^{(0)}=\left(  {%
\begin{array}
[c]{ccccc}%
-\left(  {\lambda}+{{r_{2}}}\right)   & \lambda &  &  & \\
& -\left(  {\lambda}+{{r_{2}}}\right)   & \lambda &  & \\
&  & \ddots & \ddots & \\
&  &  & -\left(  {\lambda}+{{r_{2}}}\right)   & \lambda\\
&  &  &  & -\left(  {\lambda}+{{r_{2}}}\right)
\end{array}
}\right)  ,
\]%
\[
{A_{0}}=\left(  {%
\begin{array}
[c]{cccc}%
{{r_{2}}} &  &  & \\
& {{r_{2}}} &  & \\
&  & \ddots & \\
\lambda &  &  & {{r_{2}}}%
\end{array}
}\right)  ,\text{ \ }{A_{2}}=\left(  {%
\begin{array}
[c]{cccc}%
{{r_{1}}} &  &  & \\
& {{r_{1}}} &  & \\
&  & \ddots & \\
&  &  & {{r_{1}}}%
\end{array}
}\right)  ,
\]%
\[
{A_{1}}=\left(  {%
\begin{array}
[c]{ccccc}%
-\left(  {\lambda}+{{r_{2}}+{r_{1}}}\right)   & \lambda &  &  & \\
& -\left(  {\lambda}+{{r_{2}}+{r_{1}}}\right)   & \lambda &  & \\
&  & \ddots & \ddots & \\
&  &  & -\left(  {\lambda}+{{r_{2}}+{r_{1}}}\right)   & \lambda\\
&  &  &  & -\left(  {\lambda}+{{r_{2}}+{r_{1}}}\right)
\end{array}
}\right)  .
\]
Obviously, the continuous-time Markov process $T$ is a level-independent QBD
process. Thus, we can apply the matrix-geometric solution to analyze the QBD
process $T$ and the dynamic PBFT blockchain system.

The following theorem provides a stability condition of the QBD process $T$.

\begin{The}
The level-independent QBD $T$ is positive recurrent if and only if
\[
\lambda+{r_{2}}b<{r_{1}}b.
\]
\label{the-1}
\end{The}

\textbf{Proof.} For the continuous-time QBD process $T$, we use the mean-drift
method to provide a stability condition. To use the mean-drift method, readers
may refer to Chapter 1 of Neuts \cite{Neu:1981} or Chapter 3 of Li
\cite{Li:2010}. We write
\[
\mathbf{A}={A_{2}}+{A_{1}}+{A_{0}}=\left(  {%
\begin{array}
[c]{ccccc}%
{-\lambda} & \lambda &  &  & \\
& {-\lambda} & \lambda &  & \\
&  & \ddots & \ddots & \\
&  &  & {-\lambda} & \lambda\\
\lambda &  &  &  & {-\lambda}%
\end{array}
}\right)  .
\]
Clearly, the Markov process $\mathbf{A}$ is irreducible, aperiodic and
positive recurrent. Let $\boldsymbol{\varphi} $ be the stationary probability
vector of Markov process $\mathbf{A}$, where $\boldsymbol{\varphi} =\left(
{{\varphi_{1}},{\varphi_{2}},\ldots,{\varphi_{b}}}\right)  $. Then
$\boldsymbol{\varphi} $ is the unique solution to the system of linear
equations: $\boldsymbol{\varphi} \mathbf{A}=0$ and $\boldsymbol{\varphi}
\mathbf{e}=1$. It is easy to check that ${\varphi_{1}}={\varphi_{2}}%
=\cdots={\varphi_{b}}=1/b$.

Using the mean-drift method, it is well-known that the QBD process $T$ is
positive recurrent if and only if
\[
\boldsymbol{\varphi} {A_0} \mathbf{e}<\boldsymbol{\varphi} {A_2} \mathbf{e}.
\]
Note that
\[
\boldsymbol{\varphi} {A_0} \mathbf{e}=\frac{{\lambda+{r_{2}}b}}{b}%
,\quad\boldsymbol{\varphi} {A_2} \mathbf{e}={r_{1}},
\]
this gives
\[
\lambda+{r_{2}}b<{r_{1}}b.
\]
Therefore, the QBD process $T$ is positive recurrent if and only if%
\[
\lambda+{r_{2}}b<{r_{1}}b.
\]
This completes the proof. $\square$

When the QBD process $T$ is positive recurrent, we write its stationary
probability vector as
\[
{\boldsymbol{\omega }  }=\left(  {{\omega_{0}},{\omega_{1}},{\omega_{2}%
},\ldots}\right)  ,
\]
where
\[
\omega_{k}=\left(  \omega_{kb},\omega_{kb+1},\ldots,\omega_{(k+1)b-1}\right)
,\quad k\geq0.
\]

Note that the stationary probability vector ${\boldsymbol{\omega }  }$ in
general has not an explicit expression, thus we need to develop some numerical
solution to the vector $\boldsymbol{\omega }  $. To this end, it is easy to
see from Chapter 3 of Neuts \cite{Neu:1981} that we first need to numerically
compute the rate matrix $R$, which is the minimal nonnegative solution to the
nonlinear matrix equation $R^{2}{A_{2}}+R{A_{1}}+{A_{0}}=0$. In addition, the
rate matrix $R$ can be numerically calculated by an iterative algorithm (see
Chapter 3 of Neuts \cite{Neu:1981}) as follows:
\[
R_{0}=0,
\]
and
\begin{equation}
R_{n+1}=\left(  R_{n}^{2}{A_{2}}+{A_{0}}\right)  \left(  {-{A_{1}^{-1}}%
}\right)  ,\quad n=1,2,3,\ldots. \label{equa-1}%
\end{equation}
For the matrix sequence $\left\{  R_{n},n\geq0\right\}  $, it is easy to see
that $R(n)\uparrow R$ as $n\rightarrow\infty$ by means of the Chapter 3 of
Neuts \cite{Neu:1981}, thus, for any sufficiently small positive number
$\varepsilon$, there exists a positive integer $\mathbf{n}$ such that
\begin{equation}
\left\|  R_{\mathbf{n}+1}-R_{\mathbf{n}}\right\|  <\varepsilon. \label{e18}%
\end{equation}
In this case, we take $R\approx R_{\mathbf{n}}$, which gives an approximate
solution to the nonlinear matrix equation $R^{2}{A_{2}}+R{A_{1}}+{A_{0}}=0$.

The following theorem provides expression for the stationary probability
vector ${\boldsymbol{\omega } }$, which directly comes from Theorem 1.2.1 of
Chapter 1 in Neuts \cite{Neu:1981}. Here, we restate it without a proof.

\begin{The}
If the QBD process $T$ is positive recurrent, then its stationary probability
vector ${\boldsymbol{\omega }  }=\left(  {{\omega_{0}},{\omega_{1}}%
,{\omega_{2}},\ldots}\right)  $ is given by
\begin{equation}
{\omega_{k}} = {\omega_{1}}{R^{k - 1}},k \ge1, \label{e16}%
\end{equation}
where ${\omega_{0}}$ and ${\omega_{1}}$ are the unique solution to the
following system of linear equations:
\begin{equation}
\left\{  {%
\begin{array}
[c]{l}%
{{\omega_{0}}A_{1}^{(0)} + {\omega_{1}}{A_{2}} = 0,}\\
{{\omega_{0}}{A_{0}} + {\omega_{1}}\left(  {{A_{1}} + R{A_{2}}} \right)  =
0,}\\
{{\omega_{0}}\mathbf{e}\mathrm{{ + }}{\omega_{1}}{{\left(  {I - R} \right)
}^{ - 1}}\mathbf{e} = 1.}%
\end{array}
} \right.  \label{e17}%
\end{equation}
\label{the-2}
\end{The}

\subsection{Performance analysis of the PBFT blockchain system}

Based on the $\mathrm{M}\oplus\mathrm{M}^{\mathrm{b}}/\mathrm{M}^{\mathrm{b}%
}/1$ queue and the stationary probability vector $\boldsymbol{\omega }  $, we
provide some key performance measures of the PBFT blockchain system as follows:

\textbf{(a) (i) The stationary probability of no transaction package in the
dynamic PBFT blockchain system} is given by
\begin{equation}
{\eta_{1}}={\omega_{0}}\mathbf{e}. \label{e19}%
\end{equation}
\textbf{(ii) The stationary probability of existing transaction package in the
dynamic PBFT blockchain system} is given by
\[
{\eta_{2}}=1-{\eta_{1}}=1-{\omega_{0}}\mathbf{e}.
\]

\textbf{(b) (i) The stationary rate that a block is pegged on the blockchain
in the dynamic PBFT blockchain system} is given by
\begin{equation}
{\Re_{1}}=\beta{\eta_{2}}{\zeta_{1}}={\eta_{2}}{r_{1}}. \label{e20}%
\end{equation}
\textbf{(ii) The stationary rate that an orphan block is returned to the
transaction pool} is given by
\[
{\Re_{2}}=\beta{\eta_{2}}{\zeta_{2}}={\eta_{2}}{r_{2}}.
\]

Now, we provide an effective method to compute the throughput of the dynamic
PBFT blockchain system.

\begin{The}
The transaction throughput of the dynamic PBFT blockchain system is given by
\begin{equation}
\mathrm{{TH}}={\Re_{1}}b. \label{e21}%
\end{equation}
\end{The}

\textbf{Proof.} From Figure 6, it is seen that the block throughput of the
dynamic PBFT blockchain system is given by%
\begin{align*}
\text{TH}_{\text{block}}=  &  \text{The stationary rate that a block is pegged
on the blockchain}\\
&  \times\text{The stationary probability that a block is pegged on the
blockchain, }%
\end{align*}
this gives%
\[
\text{TH}_{\text{block}}={r_{1}}\sum_{k=1}{\omega_{k}e=r_{1}}\left(
1-{\omega_{0}}\mathbf{e}\right)  ={r_{1}\eta_{2}=\Re_{1}.}%
\]
Thus, the transaction throughput of the dynamic PBFT blockchain system is
given by%
\[
\text{TH}=b\text{TH}_{\text{block}}={\Re_{1}}b.
\]
This completes the proof. $\square$

\begin{Rem}
The dynamic PBFT blockchain system is a very complicated stochastic system. To
analyze such a complicated blockchain system, this paper develops a two-stage
decomposition technique: One for the voting process corresponding to the
service times; and the other for an approximate queueing system with a
feedback mechanism. We find that the two-stage decomposition technique is very
effective for studying the PoS (or DPoS) blockchain systems, the Raft
blockchain systems and others. In particular, we provide a simple expression
for evaluating the throughput of the dynamic PBFT blockchain system.
\end{Rem}

\section{Two Algorithms}

\label{sec:algorithms} In this section, we provide two effective algorithms
through using the key techniques given in Bright and Taylor \cite{Bright:1995,
Bright:1997} and the RG-factorizations given in Li \cite{Li:2010}. In
particular, we can numerically compute the throughput of the dynamic PBFT
blockchain system.

It is worthwhile to note that the stationary rates $r_{1}$ and $r_{2}$
obtained in Section \ref{sec:performnace measures} are the elements of the
infinitesimal generator $T$ given in Section \ref{sec:dynamic system}.
Therefore, before calculating the throughput $\mathrm{{TH}}$, we need to
compute the stationary rates $r_{1}$ and $r_{2}$ firstly. To do this, we use
the RG-factorization and the method of matrix-geometric solutions given in
Neuts \cite{Neu:1981} to get the stationary rates $r_{1}$ and $r_{2}$. Such
calculation steps are shown in Algorithm \ref{alg:alg1}.

\begin{algorithm*}

  \caption{Computing the stationary rates $r_1$ and $r_2$. \label{alg:alg1}}
  \KwIn{ The key parameters: $\mu$, $\theta$, $\gamma$, $\beta$, $p$ ; \\
  \qquad\qquad Constants related to the number of nodes $L$ and $N$  }%
  \KwOut{ The stationary rates: $r_1$, $r_2$ }
 Determine transition blocks $\left\{Q_{1}^{(0)}, Q_{0}^{(0)}, Q_{2}^{(l)}, Q_{1}^{(l)}, Q_{0}^{(l)}\right\}$, $l=3L, 3L+1, \ldots, 3N+2$ \;
 Use equations (\ref{e6}-\ref{e10}) to compute U-measure, R-measure \;
 Based on the U-measure and R-measure, using $v_{0}{U_0} = 0$ and $v_{0}\mathbf{e}=1$, determine the vector $v_{0}$ \;
 Compute the stationary probabilities $\boldsymbol{\pi}_{0}$, $\boldsymbol{\pi}_{k}$, $k=3L, 3L+1, \ldots, 3N+2$ through the system of linear equations (\ref{e12}) and ${{\boldsymbol{\pi}}_{\rm{0}}}{\bf{e}} + \sum\limits_{k = 3L}^{3N + 2}{{{\boldsymbol{\pi}}_k}{\mathbf{e}}}= 1$ \;
 Use equations (\ref{e13}) and (\ref{e14}) to compute the stationary rates $r_1$ and $r_2$ \;
 Return the stationary rates $r_1$ and $r_2$.
\end{algorithm*}

Next, we use the stationary rates $r_{1}$ and $r_{2}$ obtained by Algorithm
\ref{alg:alg1} to further calculate throughput $\mathrm{{TH}}$. Note that the
calculation of throughput $\mathrm{{TH}}$ depends on $\boldsymbol{\omega } $,
and the calculation of $\boldsymbol{\omega } $ depends on the rate matrix $R$.
Therefore, we first need to determine the rate matrix $R$, and then compute
throughput $\mathrm{{TH}}$. Note that the rate matrix $R$ can be approximately
calculated by the iterative algorithm, thus, we take a controllable accuracy
$\varepsilon=10^{-12}$, and by using the equation (\ref{e18}), we can get the
rate matrix $R\approx R_{\mathbf{n}}$ and the suitable number of iterations
$\mathbf{n}$ once the termination condition is met. When we get the suitable
rate matrix $R$ and iterative number $\mathbf{n}$, we can compute the
${\eta_{1}}$ and ${\eta_{2}}$, and further we can get the approximate
throughput $\mathrm{{TH}}$ accordingly. Such calculation steps are shown in
Algorithm \ref{alg:alg2}.

\begin{algorithm*}
  \caption{Computing throughput of the dynamic PBFT blockchain system. \label{alg:alg2}}
  \KwIn{ The key parameters: $\mu$, $\theta$, $\gamma$, $\beta$, $p$, $\lambda$, $b$ ; \\
  \qquad\qquad A controllable accuracy $\varepsilon$  }%
  \KwOut{ The throughput of the dynamic PBFT blockchain system: $\rm{TH}$ }
 Use Algorithm \ref{alg:alg1}, compute the stationary rates $r_1$ and $r_2$ \;
 Based on the obtained $r_1$, $r_2$, determine the transition blocks $\left\{B_{1}^{(0)}, A_{2}, A_{1}, A_{0} \right\}$ \;
 Use equation (\ref{equa-1}) to compute $R$ iteratively, and stop the iteration if
 \[\left\|  R_{\mathbf{n}+1}-R_{\mathbf{n}}\right\|  <\varepsilon \text{ };\]  \\
 Solve $\omega _0$ and $\omega _1$ through the system equations (\ref{e17}), and then get the $\omega _k$ through equation (\ref{e16}), $k=1, 2, \ldots, K$ \;
 Compute $\Re _1$ given by equation (\ref{e20}) through equation (\ref{e19}) \;
 Compute the throughput $\rm{TH}$ through the equation (\ref{e21}) \;
 Return the throughput $\rm{TH}$.
\end{algorithm*}

\section{Numerical Analysis}

\label{sec:numberical analysis}

In this section, we use two groups of numerical examples to verify the
validity of our theoretical results and to show how some key system parameters
influence performance measures of the dynamic PBFT voting process and its
dynamic blockchain system.

\textbf{Group one: The dynamic PBFT voting process}

Now, we are going to observe the impact of the key parameters $\mu
,\theta,\gamma,p,\beta$ on the performance measures of the dynamic PBFT voting process.

\begin{figure}[ptbh]
\centering                    \subfigure[${\zeta_{1}}$, $r_{1}%
$ vs. $p$, $\mu$]{
\begin{minipage}[]{0.4\linewidth}%

\includegraphics[width=1\linewidth]{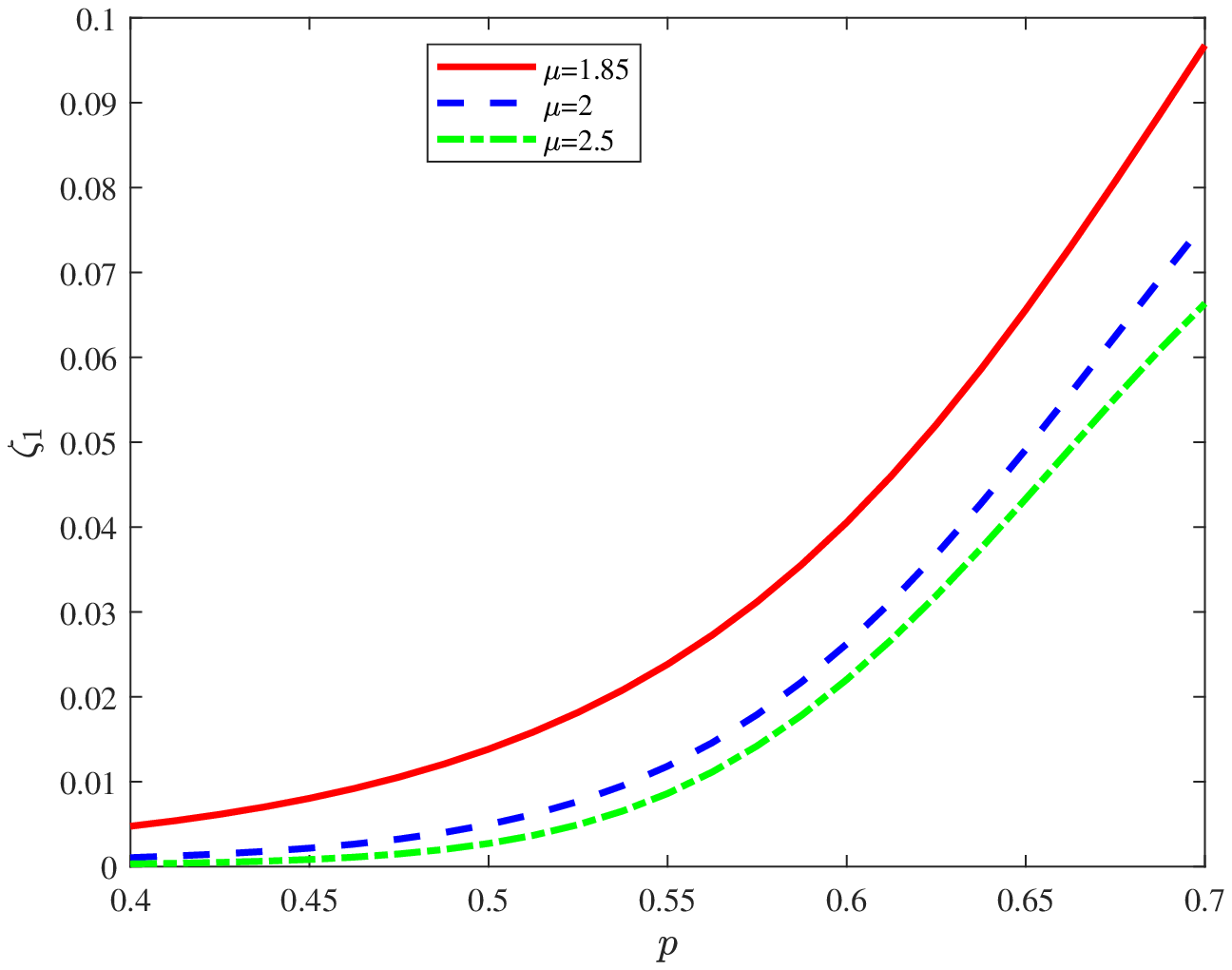}%

\
 \includegraphics[width=1\linewidth]{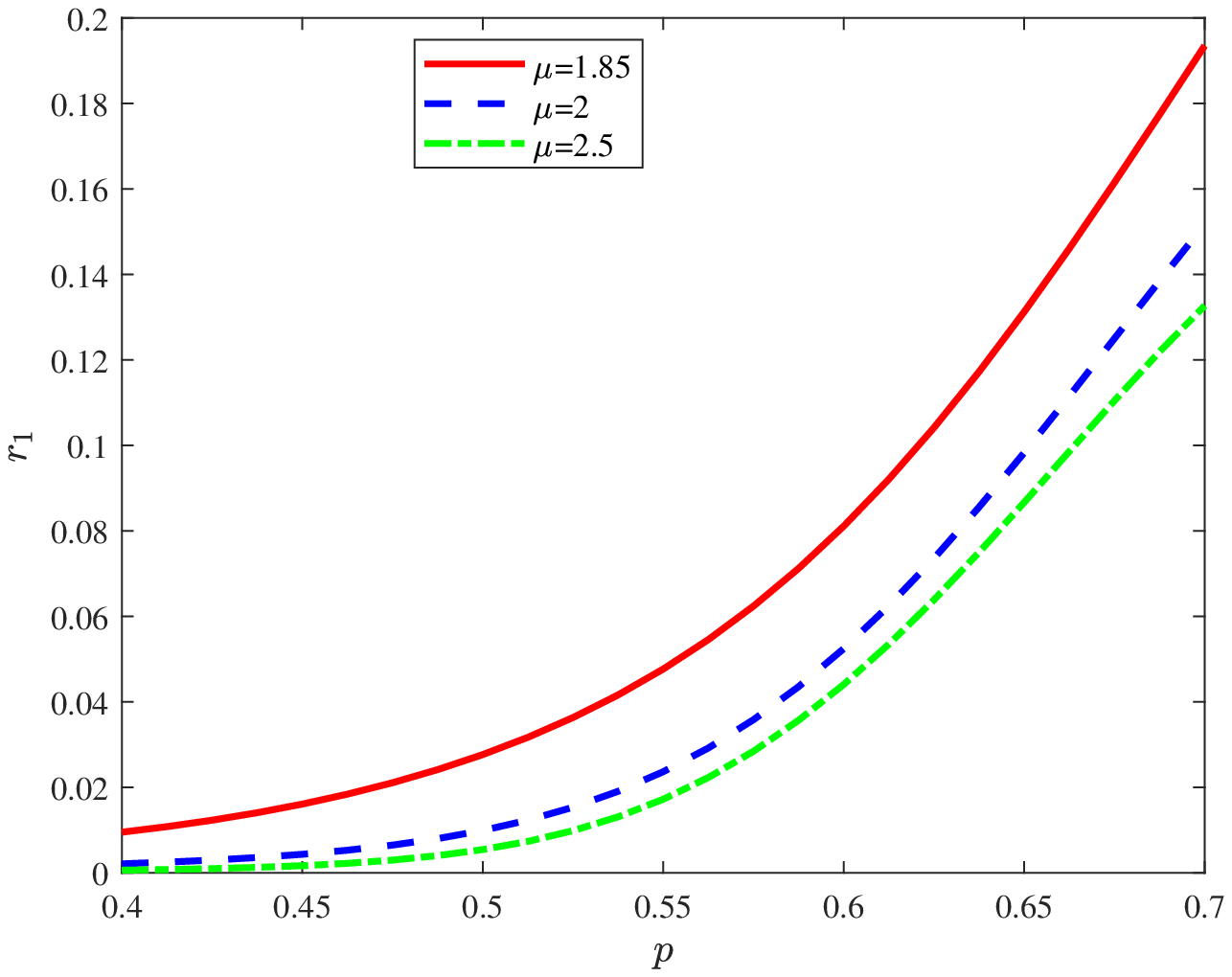}\label{figure7a}%

\end{minipage}}  \subfigure[${\zeta_{1}}$, $r_{1}%
$ vs. $p$, $\theta$]{
\begin{minipage}[]{0.4\linewidth}%

\includegraphics[width=1\linewidth]{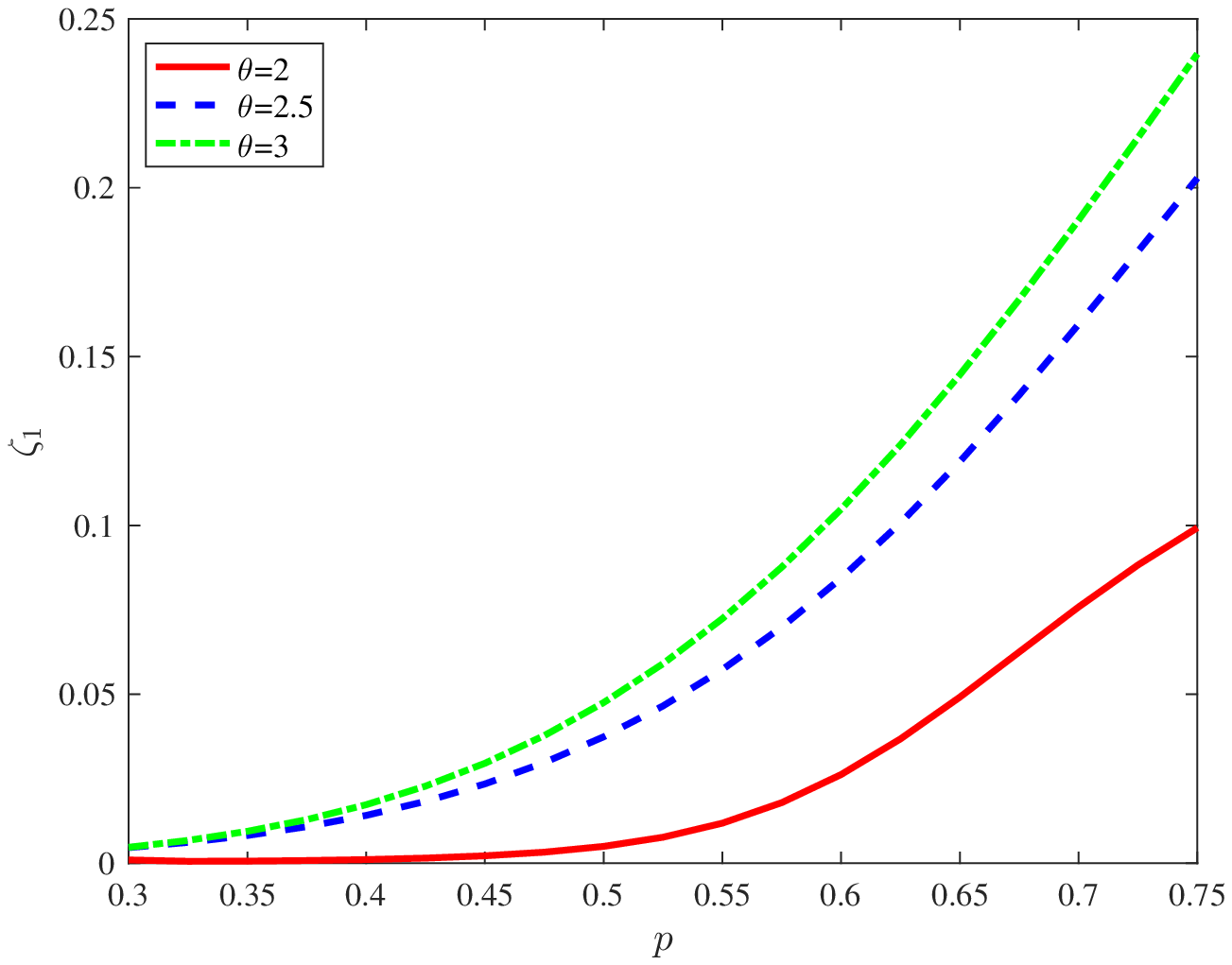}%

\
 \includegraphics[width=1\linewidth]{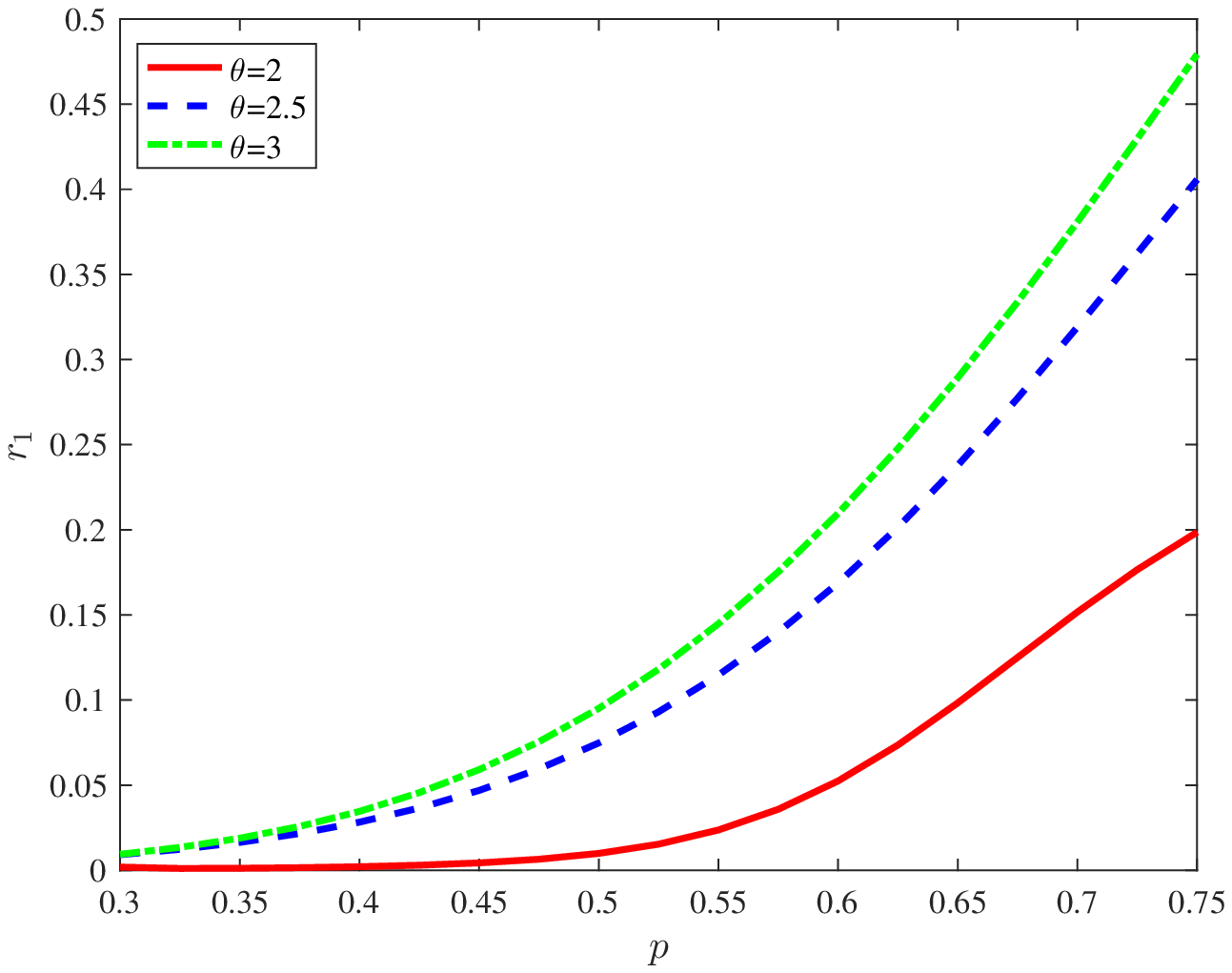}\label{figure7b}%

\end{minipage}}  \caption{Two performance measures ${\zeta_{1}}$ and $r_{1}$
vs. three parameters $p$, $\mu$, and $\theta$. }%
\label{figure:Fig-7}%
\end{figure}

In Figure \ref{figure7a}, we take the parameters as follows: $\theta=2$,
$\beta=2$, $\gamma=10$, $p\in\left[  {0.4,0.7}\right]  $ and $\mu=1.85,2,2.5$.
In Figure \ref{figure7b}, we take the parameters as follows: $\mu=2$, $\beta=
2$, $\gamma= 10$, $p \in\left[  {0.3,0.75} \right]  $, and $\theta= 2,2.5,3$.

From Figure \ref{figure:Fig-7}, it is seen that all ${\zeta_{1}}$ and $r_{1}$
increase as $p$ increases, which indicates that the stationary probability
${\zeta_{1}}$ (or rate $r_{1}$) that a transaction package becomes a block can
increase as the probability $p$ that a transaction package is approved by each
node increases. In addition, we can observe that ${\zeta_{1}}$ and $r_{1}$
decrease as $\mu$ increases in Figure \ref{figure7a}; while ${\zeta_{1}}$ and
$r_{1}$ increase as $\theta$ increases in Figure \ref{figure7b}. These
numerical results indicate that as $\mu$ increases, more and more external
nodes enter the dynamic PBFT network, such that the stationary probability
${\zeta_{1}}$ (or rate $r_{1}$) that a transaction package becomes a block can
decrease; while as $\theta$ increases, more and more external nodes leave the
dynamic PBFT network, such that the stationary probability ${\zeta_{1}}$ (or
rate $r_{1}$) that a transaction package becomes a block can increase. This
shows that the number of votable nodes in the dynamic PBFT network
significantly affects the stationary probability ${\zeta_{1}}$ (or rate
$r_{1}$) that a transaction package becomes a block. Thus, they are consistent
with our intuitive understanding.

\begin{figure}[ptbh]
\centering                    \subfigure[${\zeta_{1}}$, $r_{1}%
$ vs. $p$, $\gamma$]{
\begin{minipage}[]{0.4\linewidth}%

\includegraphics[width=1\linewidth]{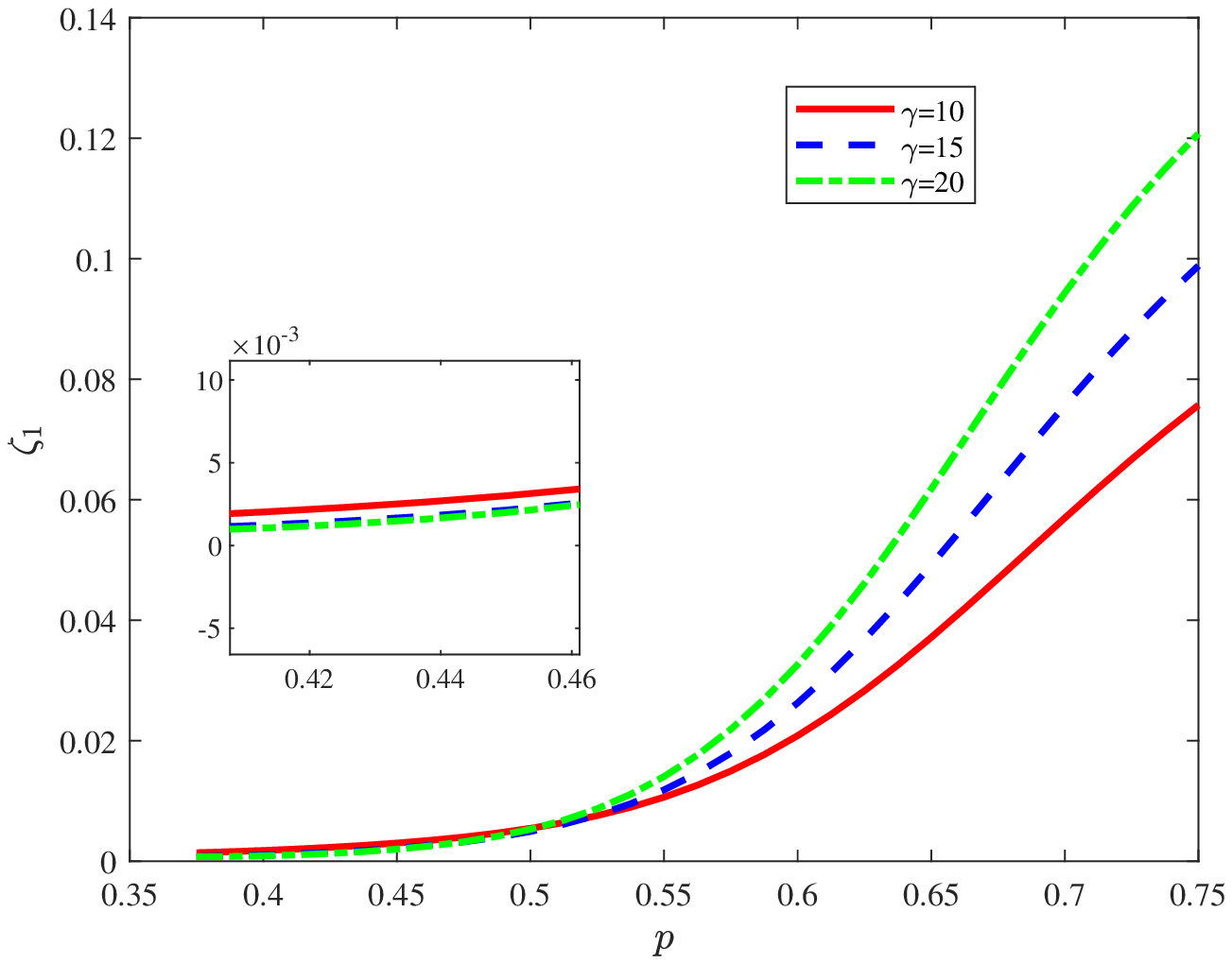}%

\
 \includegraphics[width=1\linewidth]{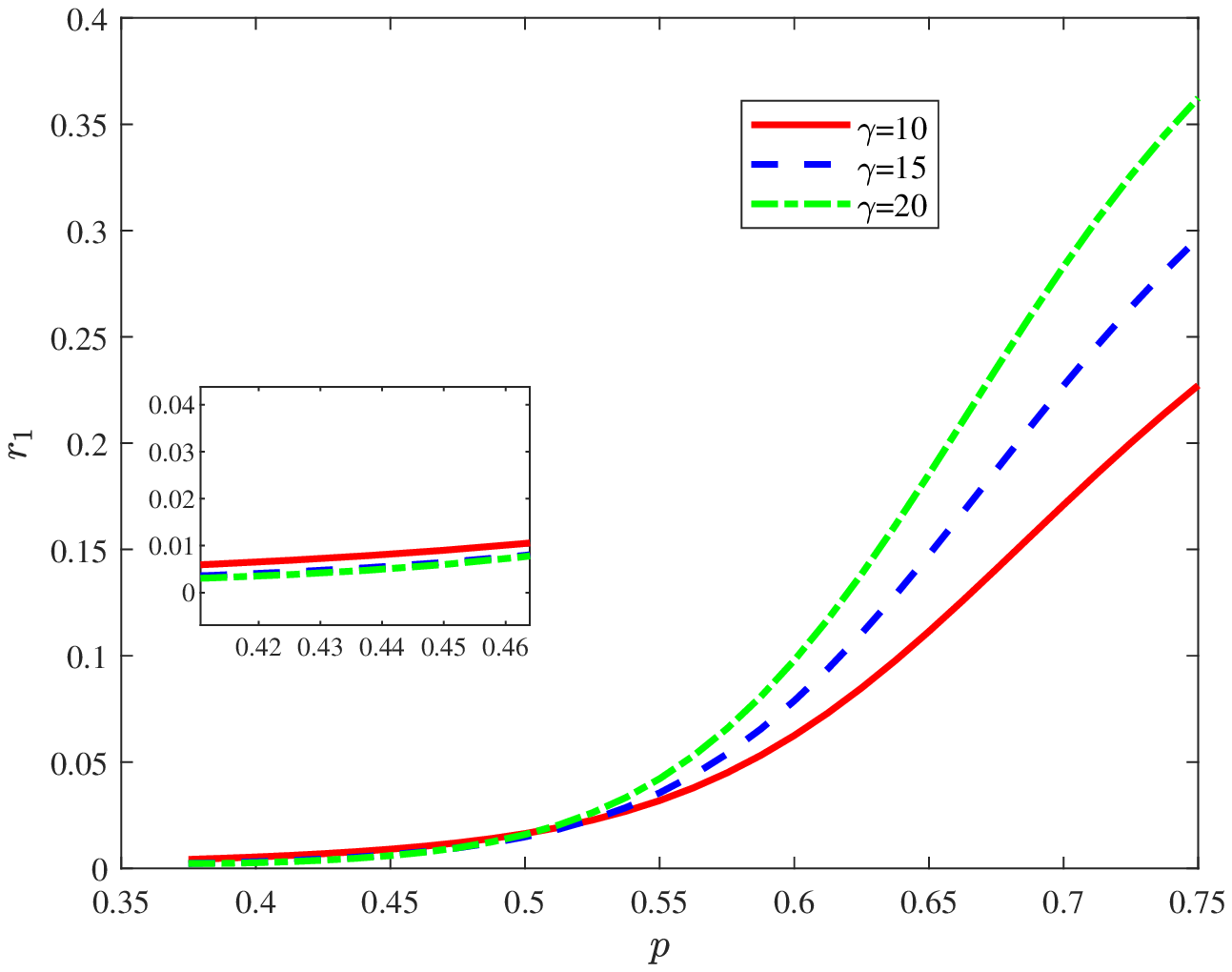}\label{figure8a}%

\end{minipage}}  \subfigure[${\zeta_{2}}$, $r_{2}%
$ vs. $p$, $\gamma$]{
\begin{minipage}[]{0.4\linewidth}%

\includegraphics[width=1\linewidth]{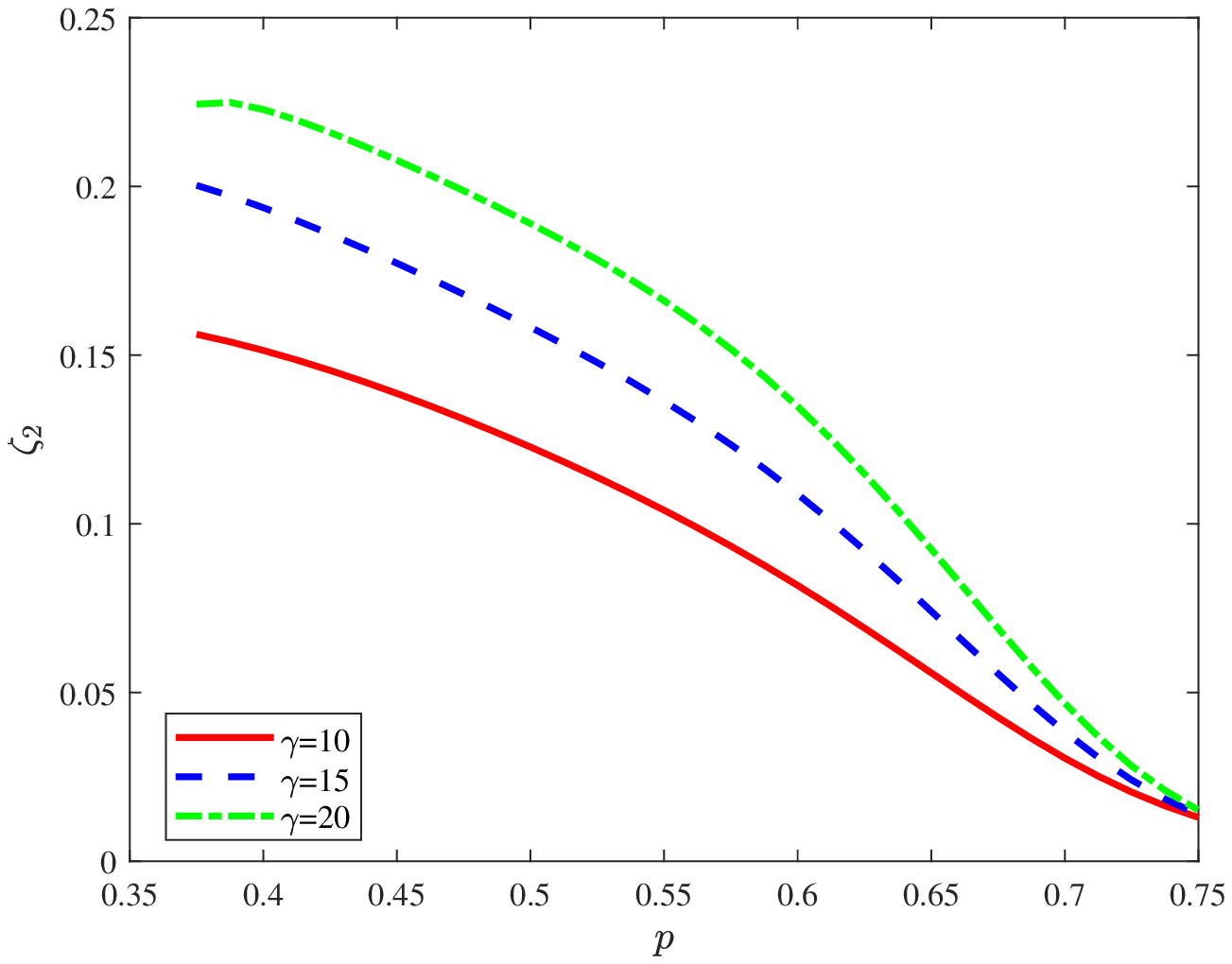}%

\
 \includegraphics[width=1\linewidth]{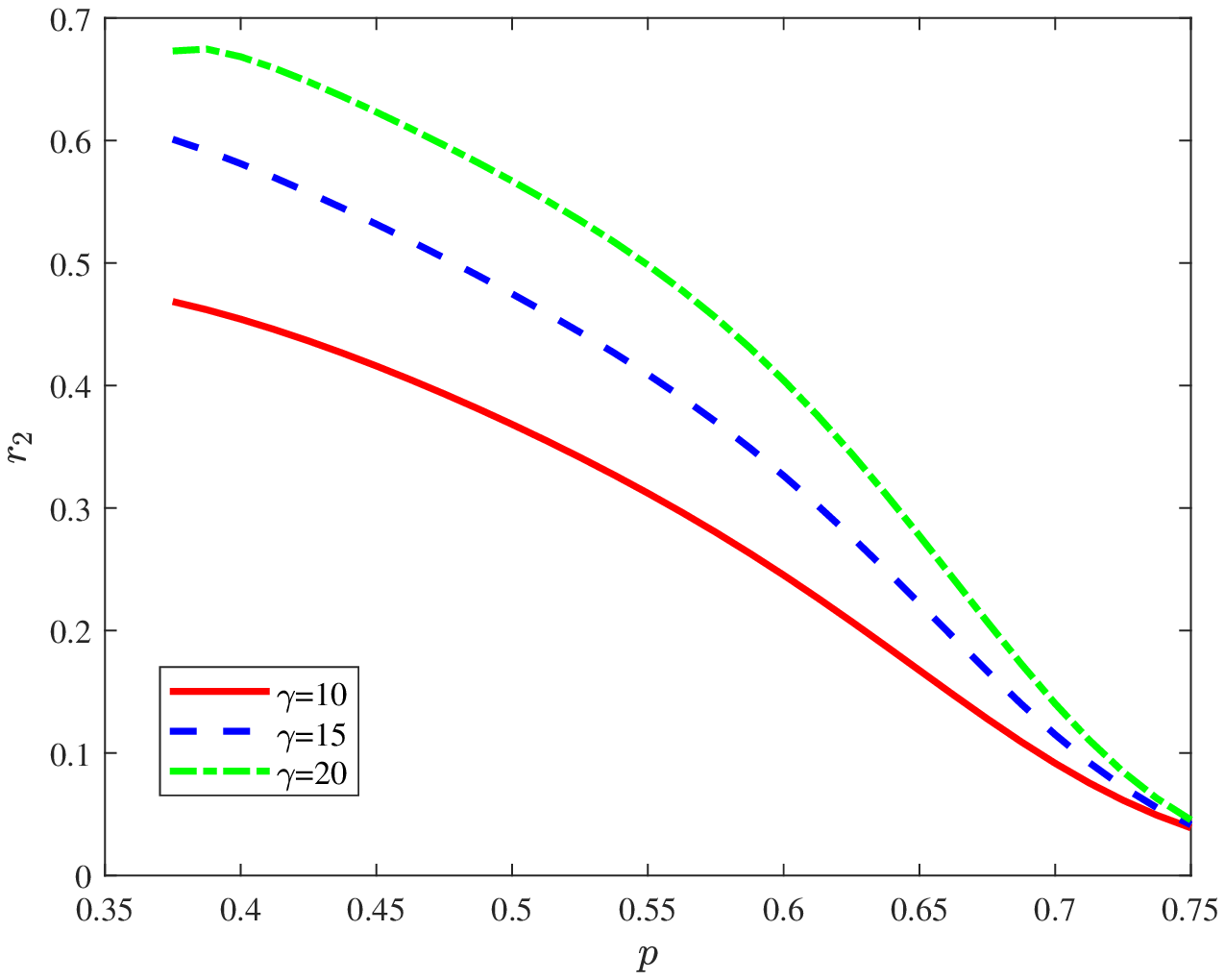}\label{figure8b}%

\end{minipage}}  \caption{Four performance measures ${\zeta_{1}}$, $r_{1}$,
${\zeta_{2}}$ and $r_{2}$ vs. two parameters $p$ and $\gamma$. }%
\label{figure:Fig-8}%
\end{figure}

In Figure \ref{figure:Fig-8}, we take the parameters as follows: $\mu=2$,
$\theta=2$, $\beta=3$, $p\in\left[  {0.375,0.75}\right]  $, and $\gamma
=10,15,20$. 

Figure \ref{figure8a} shows that ${\zeta_{1}}$ and $r_{1}$ increase as $p$
increases, sharing the same trends as that in Figure \ref{figure:Fig-7}; while
Figure \ref{figure8b} shows that ${\zeta_{2}}$ and $r_{2}$ decrease as $p$
increases. These numerical results indicate that the stationary probability
${\zeta_{1}}$ (or rate $r_{1}$) that a transaction package becomes a block
increases as the probability $p$ that a package is approved by each node
increases; while the stationary probability ${\zeta_{2}}$ (or rate $r_{2}$)
that a transaction package becomes an orphan block decreases as the
probability $p$ that a package is approved by each node increases. At the same
time, from Figure \ref{figure8a}, we can see that there exists a ${p_{0}}$
such that ${\zeta_{1}}$ and $r_{1}$ decrease as $\gamma$ increases when
$p<{p_{0}}$. This shows that the faster the votable nodes vote, the lower the
stationary probability ${\zeta_{1}}$ (or rate $r_{1}$) that the transaction
package is approved as a block. On the other hand, ${\zeta_{1}}$ and $r_{1}$
increase as $\gamma$ increases when $p>{p_{0}}$. This shows that the faster
the votable nodes vote, the greater the stationary probability ${\zeta_{1}}$
(or rate $r_{1}$) that the transaction package is approved as a block. In
addition, as can be seen from Figure \ref{figure8b}, ${\zeta_{2}}$ and $r_{2}$
increase as $\gamma$ increases. This shows that the faster the votable nodes
vote, the greater the stationary probability ${\zeta_{1}}$ (or rate $r_{1}$)
that the transaction package is refused as an orphan block. These numerical
results are also in line with our intuitive understanding.

\begin{figure}[ptbh]
\centering         \subfigure[]  {\includegraphics[width=7cm]{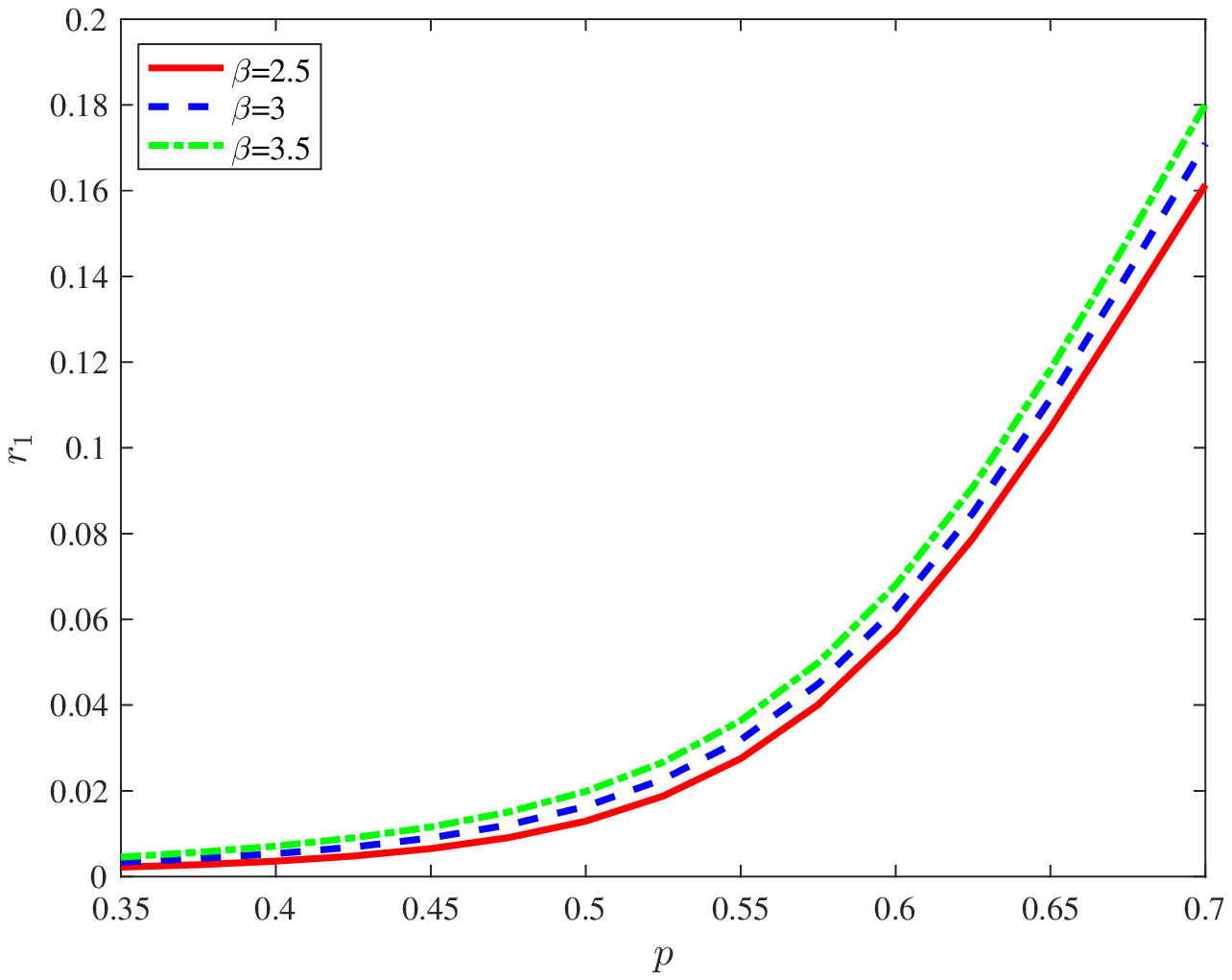}%

\label{figure9a}
}  \subfigure[]  { \includegraphics[width=7cm]{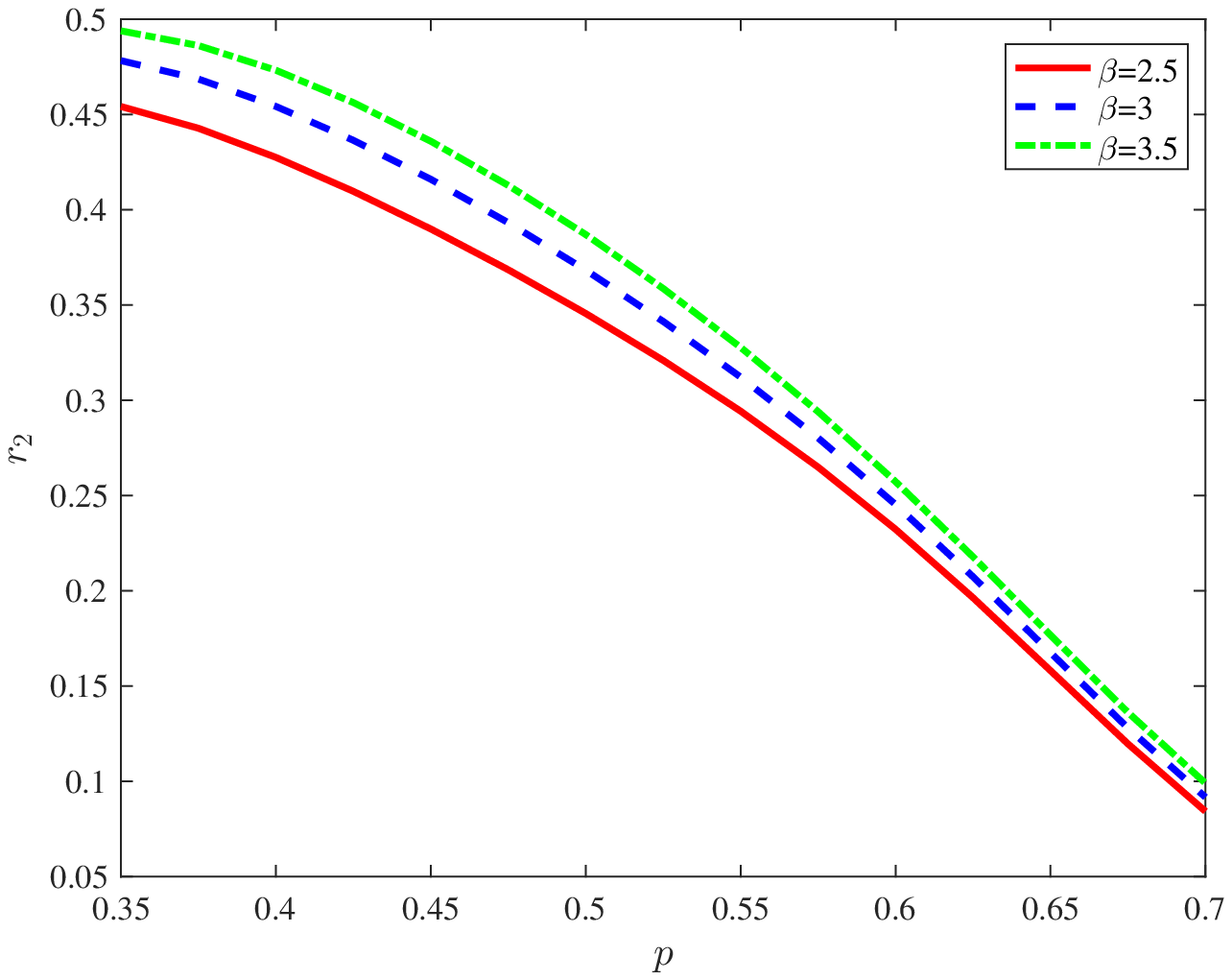}%

\label{figure9b} }  \caption{Two performance measures $r_{1}$ and ${r_{2}}$
vs. two parameters $p$ and $\beta$.}%
\label{figure:Fig-9}%
\end{figure}

In Figure \ref{figure:Fig-9}, we take the parameters as follows: $\mu=2$,
$\theta= 2$, $\gamma= 10$, $p\in\left[  {0.35,0.7} \right]  $, and $\beta=
2.5,3,3.5$.

As shown in Figure \ref{figure9a}, $r_{1}$ increases as $p$ increases; while
Figure \ref{figure9b} suggests $r_{2}$ decreases as $p$ increases. At the same
time, we can see that $r_{1}$ and $r_{2}$ increase as $\beta$ increases, which
indicates that the lower the rate $\beta$ that the network latency of the
dynamic PBFT blockchain system, the faster the rate $r_{1}$ or $r_{2}$ that
the transaction packages are pegged on the blockchain or are returned to the
transaction pool. Such a numerical result is consistent with our intuitive understanding.

\textbf{Group two: The dynamic PBFT blockchain system}

We observe the impact of $\lambda, b, \mu,\theta,\gamma,p,\beta$ on the
performance measures of the dynamic PBFT blockchain system.

Firstly, we explore the impact of $\lambda$ and $b$ on the ${\eta_{1}}$,
${\eta_{2}}$ and $\mathrm{{TH }}$. To this end, we take the some parameters as
follows: $r_{1}=0.7$, $r_{2}=0.1$, $b \in\left[  {50,250} \right]  $, and
$\lambda= 1,5,9$.

\begin{figure}[ptbh]
\centering                    \subfigure[]{
\begin{minipage}[]{0.4\linewidth}%

\includegraphics[width=1\linewidth]{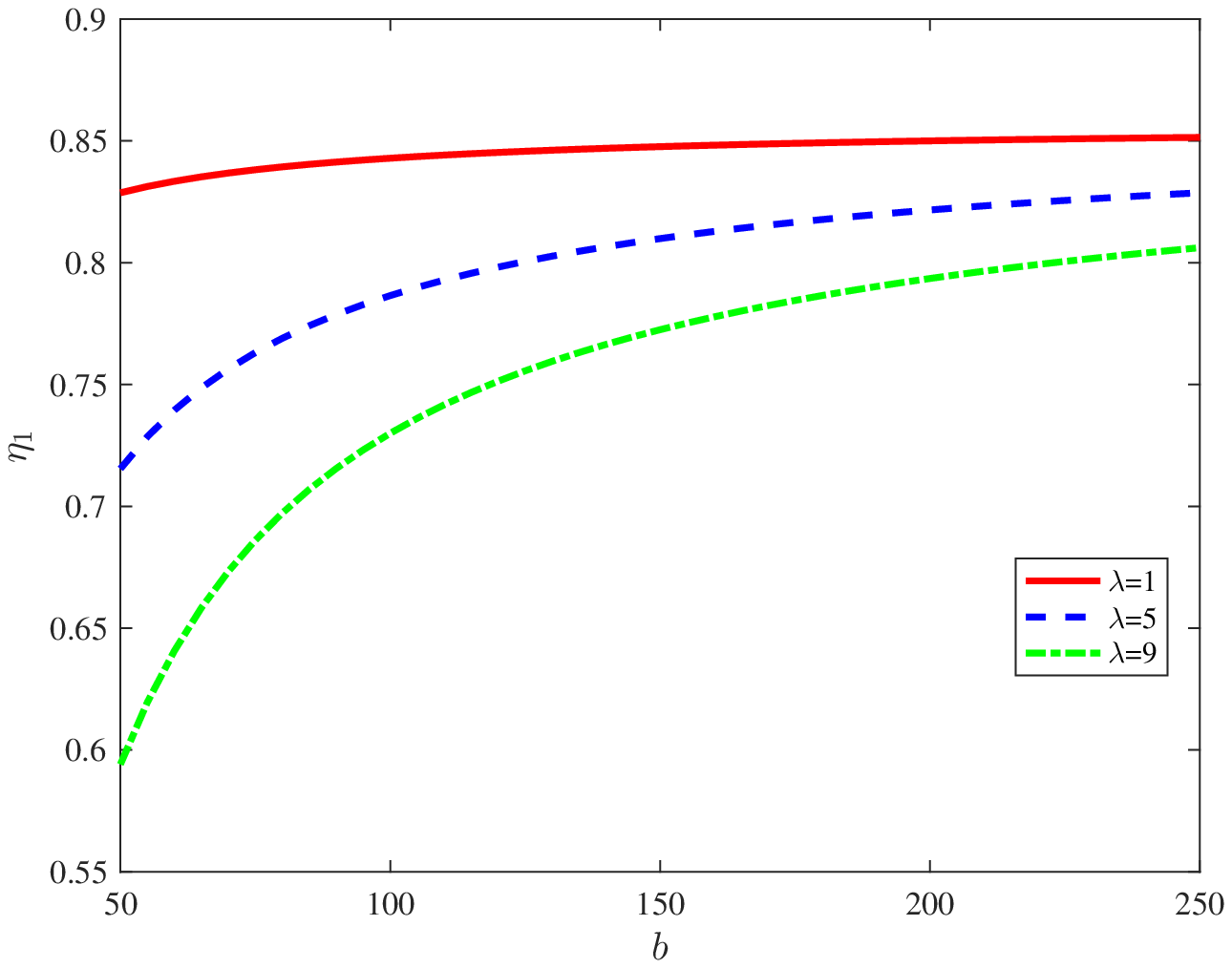}\label{figure10a}%

\end{minipage}}  \subfigure[]{
\begin{minipage}[]{0.4\linewidth}%

\includegraphics[width=1\linewidth]{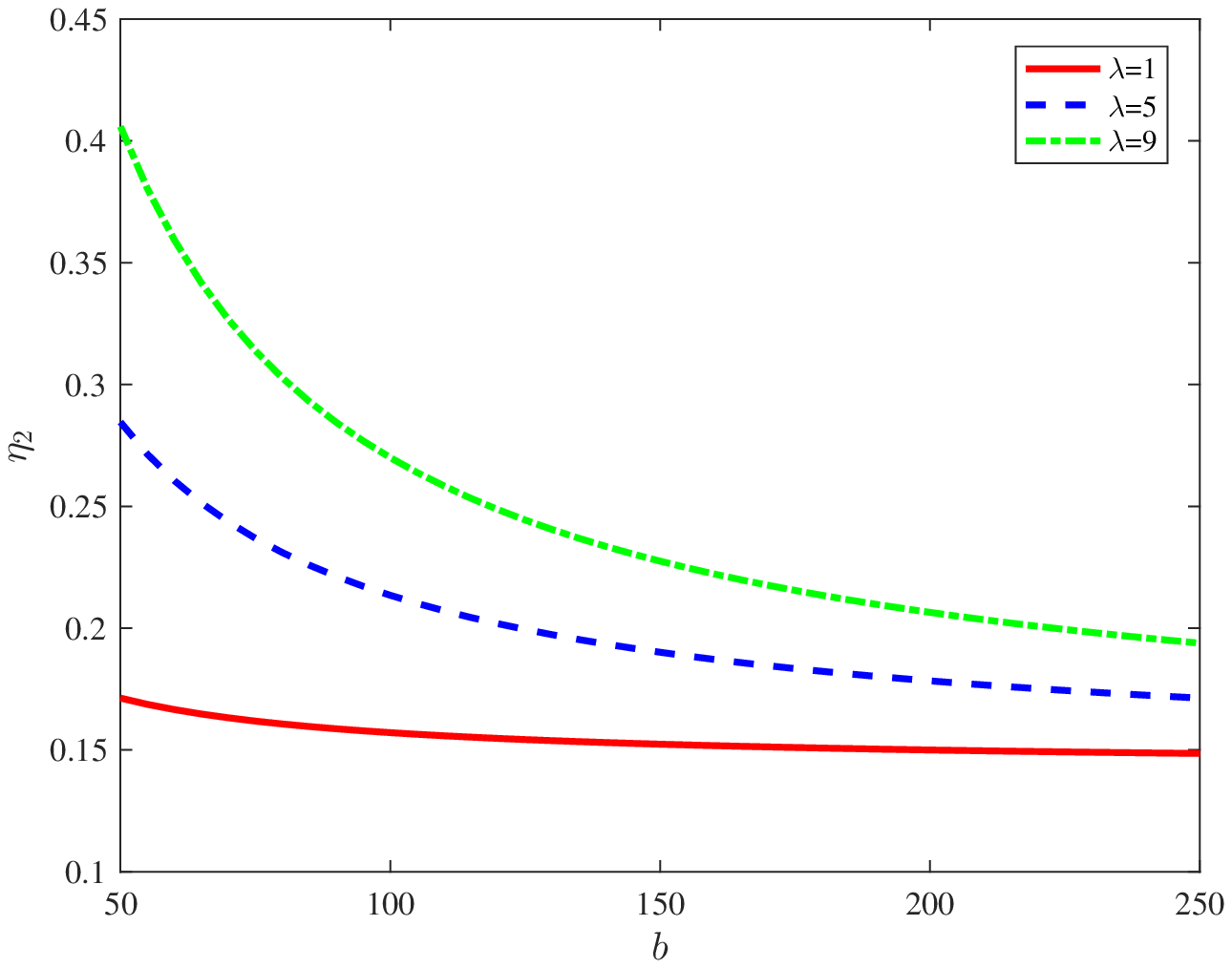}\label{figure10b}%

\end{minipage}}  \subfigure[]{
\begin{minipage}[]{0.4\linewidth}%

\includegraphics[width=1\linewidth]{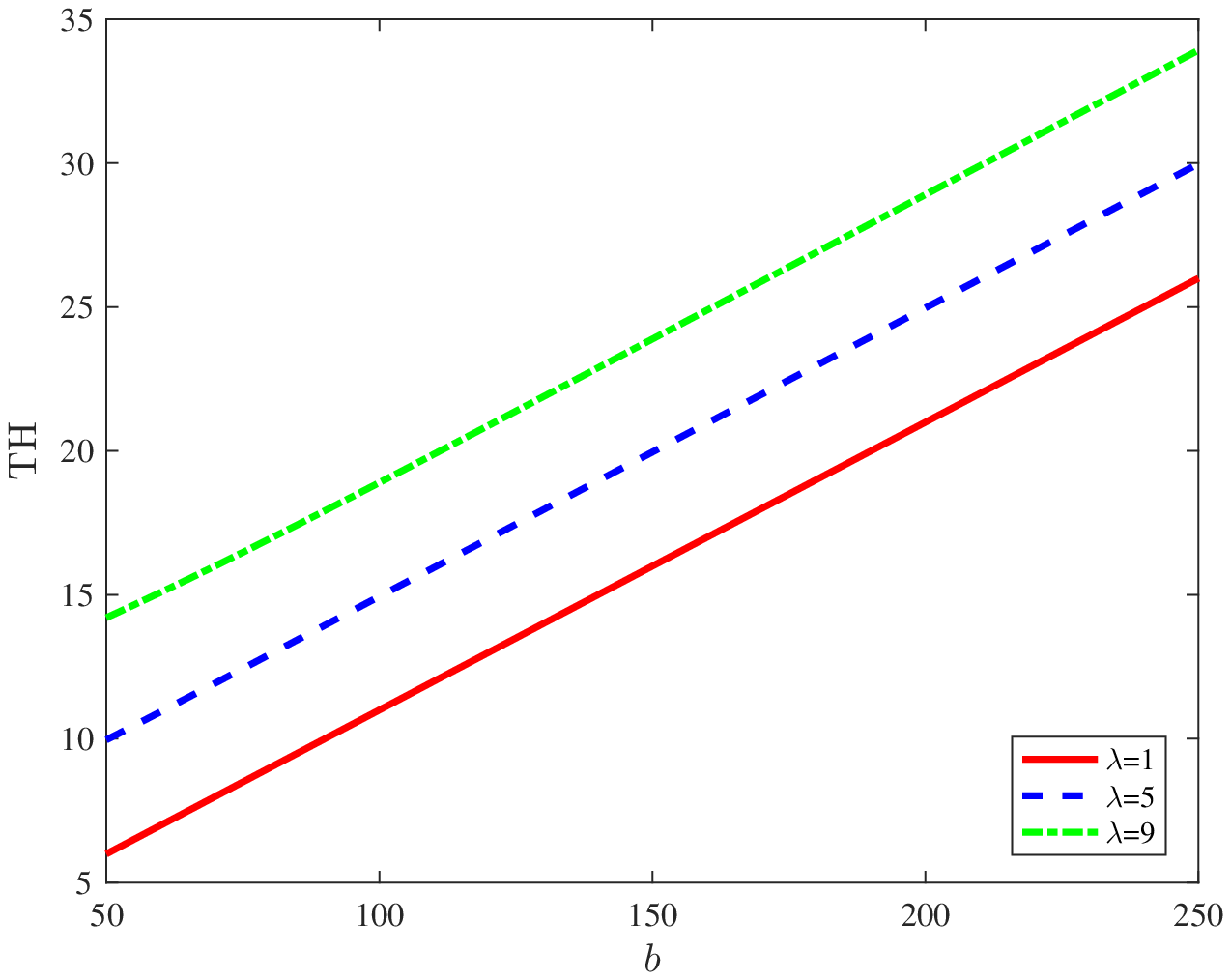}\label{figure10c}%

\end{minipage}}  \caption{Three performance measures ${\eta_{1}}$,
${\eta_{2}}$ and $\mathrm{{TH }}$ vs. two parameters $b$ and $\lambda$.}%
\label{figure:Fig-10}%
\end{figure}

From Figure \ref{figure:Fig-10}, we can see that ${\eta_{1}}$ in Figure
\ref{figure10a} and $\mathrm{{TH}}$ in Figure \ref{figure10c} increase as $b$
increases; while ${\eta_{2}}$ in Figure \ref{figure10c} decreases as $b$
increases. Such numerical results indicate that the larger the batch size $b$
is, the greater the probability ${\eta_{1}}$ of no transaction package, the
smaller the probability ${\eta_{2}}$ of the existing transaction package, and
the greater the throughput $\mathrm{{TH}}$ of the dynamic PBFT blockchain
system. In other words, as $b$ increases, the probability that the dynamic
PBFT blockchain system is in the idle period increases; while the probability
that the dynamic PBFT blockchain system is in the busy period decreases;
however, the throughput of the dynamic PBFT blockchain system does not
decrease accordingly.

Meanwhile, ${\eta_{1}}$ decreases as $\lambda$ increases, while ${\eta_{2}}$
and $\mathrm{{TH }}$ increase as $\lambda$ increases. This indicates that as
$\lambda$ increases, more and more transactions arrive in the dynamic PBFT
blockchain system, this decreases the probability ${\eta_{1}}$ of no
transaction package, increases the probability ${\eta_{2}}$ of existing
transaction package, and improves the throughput $\mathrm{{TH }}$ of the
dynamic PBFT blockchain system. Such numerical results are also consistent
with our intuitive understanding.

Secondly, we show the impact of $\mu,\theta,\gamma,p,\beta$ on $\mathrm{{TH }%
}$. To this end, we take $\lambda=10$, $b=150$ for all the following numerical examples.

\begin{figure}[ptbh]
\centering                    \subfigure[$\rm{TH }$ vs. $p,\mu$.]
{\includegraphics[width=7cm]{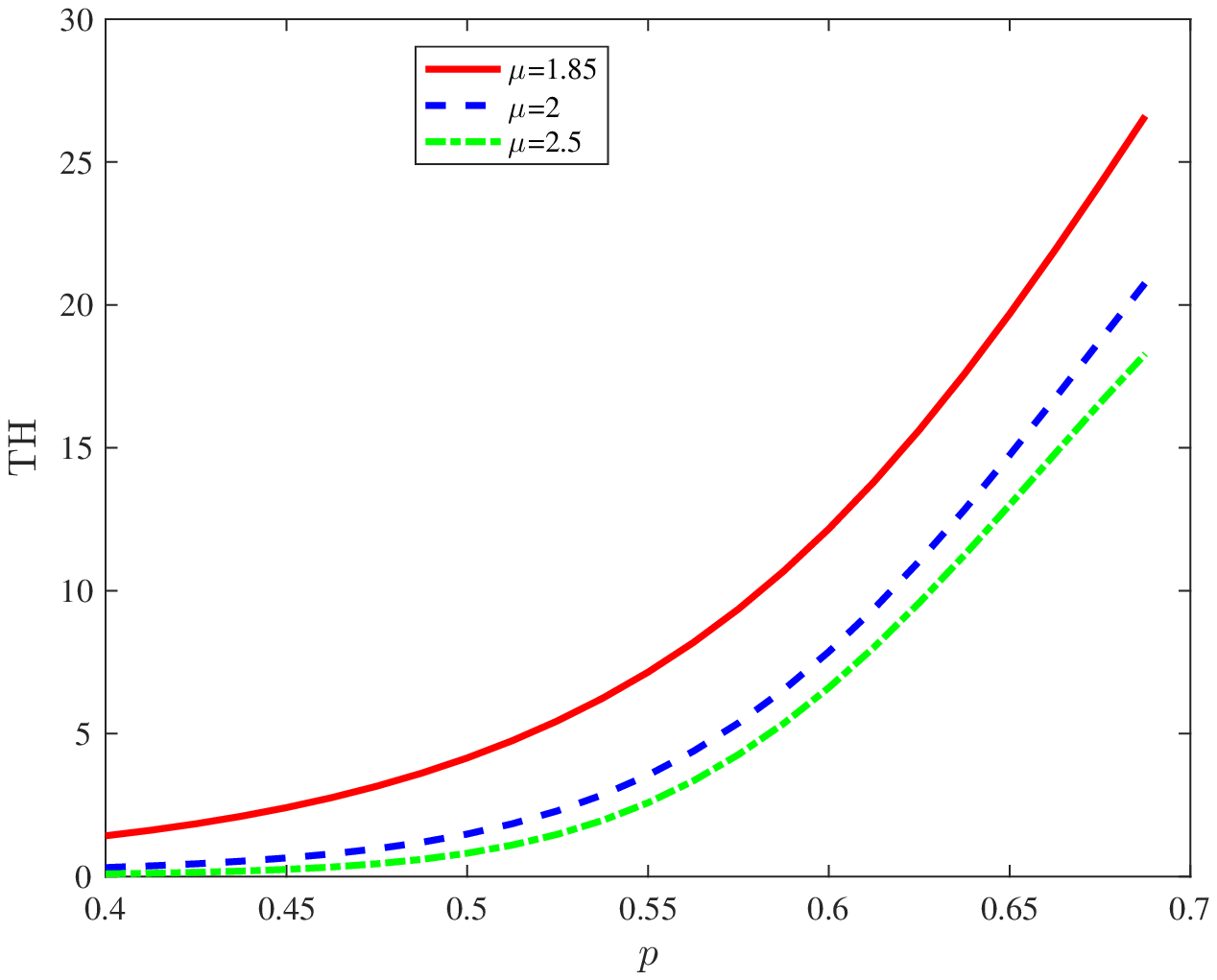}  \label{figure11a}}
\subfigure[$\rm{TH }%
$ vs. $p,\theta$.]  {
\includegraphics[width=7cm]{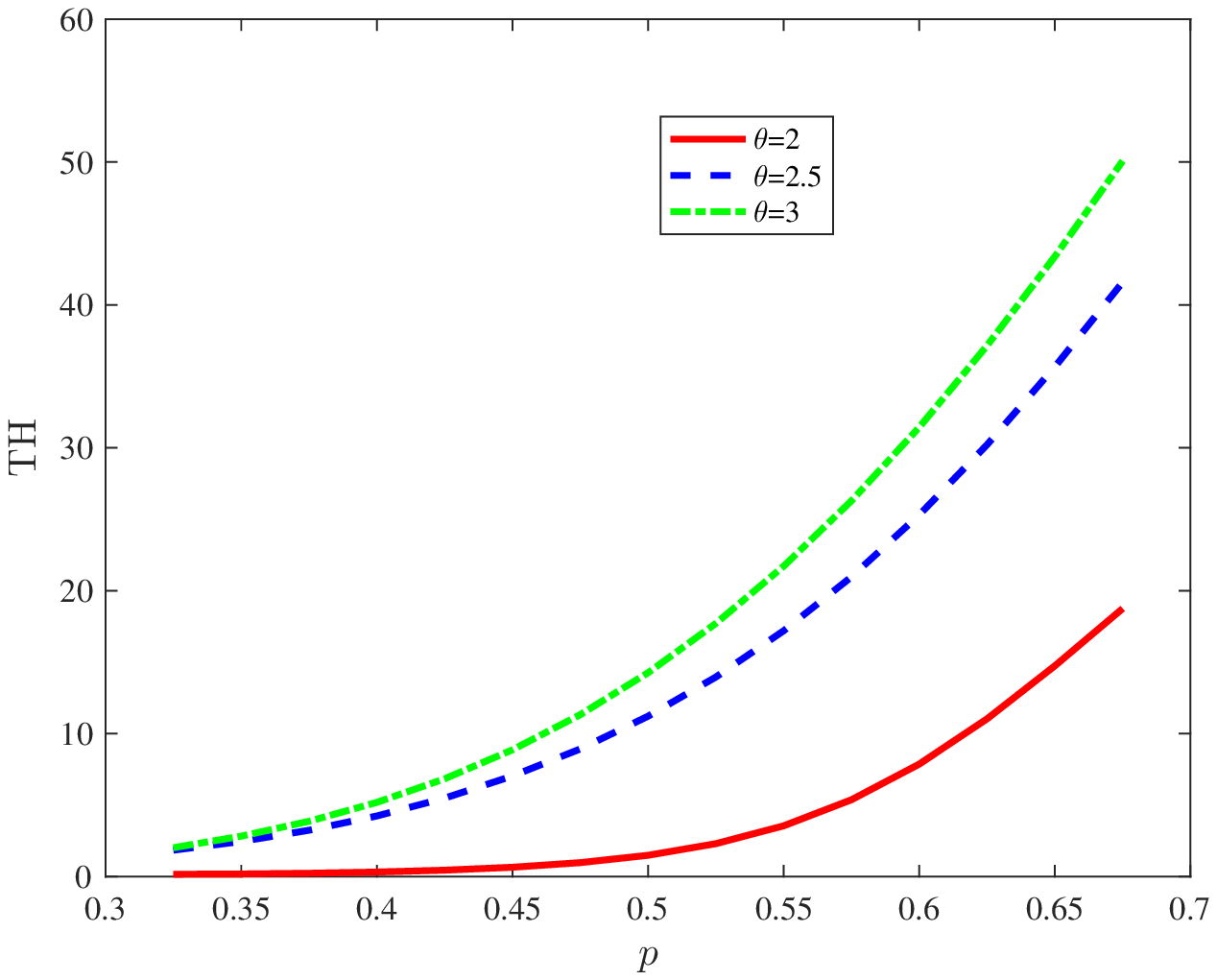}%

\label{figure11b}}  \centering                   \caption{$\mathrm{{TH }}$
vs. $p$, $\mu$, $\theta$.}%
\label{figure:Fig-11}%
\end{figure}

In Figure \ref{figure11a}, we take the parameters: $\theta= 2$, $\beta= 2$,
$\gamma= 10$, $p \in\left[  {0.4,0.6875} \right]  $, and $\mu= 1.85,2,2.5$. In
Figure \ref{figure11b}, we take the parameters: $\mu=2$, $\beta= 2$, $\gamma=
10$, $p \in\left[  {0.325,0.675} \right]  $, and $\theta= 2.5,3,3.5$.

From Figure \ref{figure11a}, we can see that $\mathrm{{TH}}$ decreases as
$\mu$ increases; while from Figure \ref{figure11b}, we can see that
$\mathrm{{TH}}$ increases as $\theta$ increases. These findings indicates that
the faster the nodes enter the dynamic PBFT network, the lower the transaction
throughput $\mathrm{{TH}}$ of the dynamic PBFT blockchain system; while the
faster the nodes leave the dynamic PBFT blockchain system, the greater the
transaction throughput $\mathrm{{TH}}$ of the dynamic PBFT blockchain system.
In addition, the number of votable nodes affects the throughput of the dynamic
PBFT blockchain system. Here, we can get a case: If we aim to pursue the high
throughput of the dynamic PBFT blockchain system, we need a small number of
votable nodes, but if most of these nodes are Byzantine, the dynamic PBFT
blockchain system will be insecure, which means that we sometimes have to
sacrifice the throughput to keep the dynamic PBFT blockchain system secure. In
addition, from Figure \ref{figure:Fig-11}, we can see that $\mathrm{{TH}}$
increases as $p$ increases, this is consistent with our intuitive understanding.

\begin{figure}[ptbh]
\centering                    \includegraphics[width=12cm]{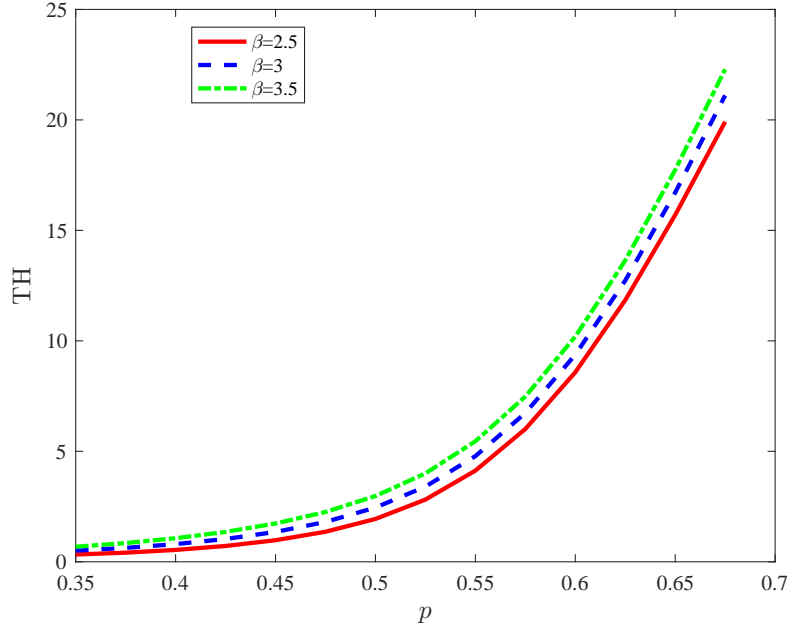}
\centering         \caption{$\mathrm{{TH }}$ vs. $p$ and $\beta$.}%
\label{figure:Fig-12}%
\end{figure}

In Figure \ref{figure:Fig-12}, we take the parameters: $\mu=2$, $\theta=2$,
$\gamma=10$, $p\in\left[  {0.3,0.6875}\right]  $, and $\beta=2.75,3,3.5$. From
Figure \ref{figure:Fig-12}, we can see that $\mathrm{{TH}}$ increases as
$\beta$ increases. This means that the faster the rate of the block-pegging or
rolling-back, the greater the transaction throughput $\mathrm{{TH}}$ of the
dynamic PBFT blockchain system. Such a numerical result is consistent with our
intuitive understanding. Meanwhile, from Figure \ref{figure:Fig-12}, we can
see that $\mathrm{{TH}}$ increases as $p$ increases, which has the same trend
as that in Figure \ref{figure:Fig-11}.

\begin{figure}[ptbh]
\centering                    \includegraphics[width=12cm]{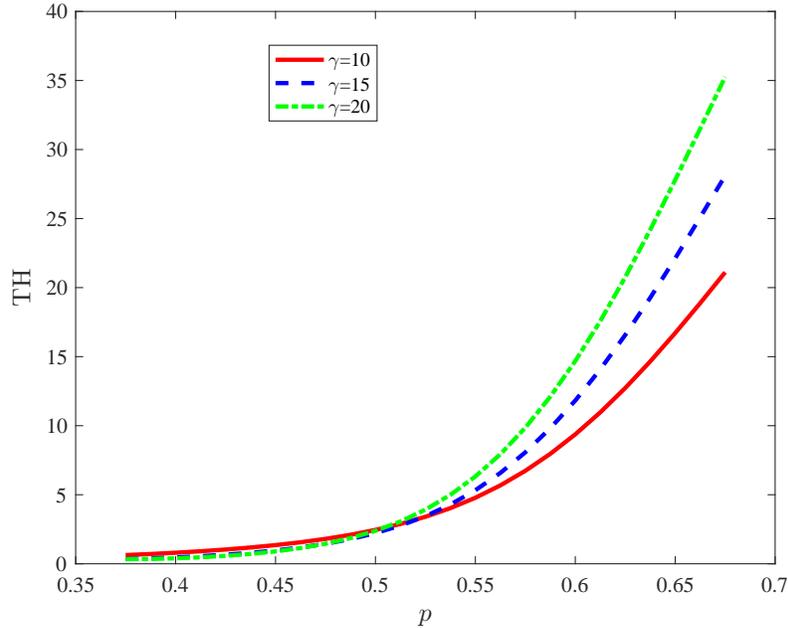}
\centering          \caption{$\mathrm{{TH }}$ vs. $p$ and $\gamma$.}%
\label{figure:Fig-13}%
\end{figure}

In Figure \ref{figure:Fig-13}, we take the parameters: $\mu=2$, $\theta=2$,
$\beta=3$, $p\in\left[  {0.325,0.675}\right]  $, and $\gamma=10,15,20$. From
Figure \ref{figure:Fig-13}, we can see that there exists a ${\tilde{p}_{0}}$
such that $\mathrm{{TH}}$ decreases as $\gamma$ increases when $p<{\tilde
{p}_{0}}$. This means that the faster the rate of votable nodes voting, the
lower the transaction throughput $\mathrm{{TH}}$ of the dynamic PBFT
blockchain system. While $\mathrm{{TH}}$ increases as $\gamma$ increases when
$p>{\tilde{p}_{0}}$. This means that the faster the rate of node voting, the
greater the transaction throughput $\mathrm{{TH}}$ of the dynamic PBFT
blockchain system. These numerical results further validate the trends of
${\zeta_{1}}$ and $r_{1}$ in Figure \ref{figure8a}. Also, from Figure
\ref{figure:Fig-13}, we can see that $\mathrm{{TH}}$ increases as $p$
increases, which has the same trend as that in Figure \ref{figure:Fig-11} as well.

\section{Concluding Remarks}

\label{sec:concluding}

In this paper, we first propose a new dynamic PBFT to generalize the ordinary
PBFT by introducing new dynamic nodes. That is, the votable nodes may always
leave the PBFT network while some new nodes can also enter the PBFT network.
Therefore, the number of votable nodes is constantly changing. Then we provide
a large-scale Markov modeling technique to analyze the dynamic PBFT voting
processes and the dynamic PBFT blockchain system. To this end, we set up a
large-scale Markov process and provide key performance analysis for both the
dynamic PBFT voting processes and the dynamic PBFT blockchain system. In
particular, we provide two effective algorithms for computing the throughput
of the dynamic PBFT blockchain system. Finally, we use numerical examples to
check the validity of our theoretical results and indicate how some key system
parameters influence the performance measures of the dynamic PBFT voting
processes and of the dynamic PBFT blockchain system.

Using the theory of multi-dimensional Markov processes and the
RG-factorization technique, we are optimistic that the methodology and results
developed in this paper shed light on the study of dynamic PBFT blockchain
systems such that a series of promising research can be developed potentially.
Along this line, we will continue our future research on several interesting
directions as follows:

--- Let all the three stages (\textit{prepare}, \textit{commit,}
and\ \textit{reply}) follow different exponential distributions with rates
$\mu_{1}$, $\mu_{2}$, and $\mu_{3}$, respectively. Note that such a
generalization is far more difficult than this paper due to some complicated
parallel phase-type calculations.

--- When the arrivals of new nodes or the departures of votable nodes are a
Markovian arrival process (MAP), an interesting future research is to focus on
finding effective algorithms for dealing with the multi-dimensional Markov
processes with a block structure corresponding to the dynamic PBFT blockchain systems.

--- When the arrivals of new nodes or the departures of votable nodes are a
renewal process, an interesting future research is to focus on fluid and
diffusion approximations of the dynamic PBFT blockchain systems.

--- Setting up reward functions with respect to cost structure, transaction
fee, block reward, blockchain security and so forth. It is very interesting in
our future study to develop stochastic optimization, Markov decision processes
and stochastic game models in the study of dynamic PBFT blockchain systems.

\section*{Acknowledgment}

Quan-Lin Li was supported by the National Natural Science Foundation of China
under grants No. 71671158 and 71932002.

\end{document}